\documentclass[letterpaper]{article} 
\usepackage{aaai2027}  
\usepackage[hyphens]{url}  
\usepackage{graphicx} 
\usepackage{natbib}  
\usepackage{caption} 
\usepackage{algorithm}
\usepackage{algorithmic}

\usepackage{enumitem}
\usepackage{multirow}
\usepackage{amssymb}
\usepackage{colortbl}

\newcommand{\tool}{{VCE-Skill}}

\usepackage{newfloat}
\usepackage{listings}
\DeclareCaptionStyle{ruled}{labelfont=normalfont,labelsep=colon,strut=off} 
\floatstyle{ruled}
\newfloat{listing}{tb}{lst}{}
\floatname{listing}{Listing}

\usepackage{booktabs}

\title{{\tool}: Enhancing Skill Self-Evolution with Version-Change Experience}
\author{
    Jianming Chen\textsuperscript{\rm 1,\rm 2,\rm 3,\rm 4},
    Xuanbin Ye\textsuperscript{\rm 5},
    Yawen Wang\textsuperscript{\rm 1,\rm 2,\rm 3,\rm 4}\thanks{Corresponding authors.},
    Junjie Wang\textsuperscript{\rm 1,\rm 2,\rm 3,\rm 4}\footnotemark[1],
    Yuanzhe Hu\textsuperscript{\rm 1,\rm 2,\rm 3,\rm 4},\\
    Qing Wang\textsuperscript{\rm 1,\rm 2,\rm 3,\rm 4},
    Fanjiang Xu\textsuperscript{\rm 1,\rm 2,\rm 3,\rm 4}\footnotemark[1]
}
\affiliations {
    \textsuperscript{\rm 1}
    Institute of Software Chinese Academy of Sciences, Beijing, China
    \\
    \textsuperscript{\rm 2}
    Science \& Technology on Integrated Information System Laboratory, Beijing, China
    \\
    \textsuperscript{\rm 3}
    State Key Laboratory of Complex System Modeling and Simulation Technology, Beijing, China
    \\
    \textsuperscript{\rm 4}
    University of Chinese Academy of Sciences, Beijing, China
    \\
    \textsuperscript{\rm 5}
    Beijing University of Post and Telecommunications, Beijing, China
}

\begin{document}

\maketitle

\begin{abstract}
Agents increasingly rely on reusable skills to encode task knowledge, tool-use procedures, and validation rules. 
Existing skill self-evolution methods primarily revise skills using execution trajectories collected from current tasks, leaving the evolution knowledge accumulated in public skill version histories largely untapped. 
Our pilot study reveals a clear complementarity between the two sources: public skill changes provide reusable evolution priors, whereas trajectories provide evidence grounded in the current task. 
Motivated by this, we propose \tool{}, which distills noisy and implementation-specific public skill changes into reusable, structured version-change experience and adaptively fuses it with trajectory-derived proposals from the base evolver, thereby exploiting external experience while retaining task-specific evidence.
Extensive experiments demonstrate that \tool{} improves skill self-evolution, increasing mean scores by 3.20--4.98 points; transfer experiments further show that the resulting skills achieve stronger cross-model transfer performance. 
Our work highlights public skill version changes as a previously underexplored yet effective source of prior knowledge and advances trajectory-driven skill self-evolution.
\end{abstract}

\section{Introduction}
\label{sec:intro}

Large language model agents increasingly rely on external skills as reusable units of task knowledge \cite{SkillsBench,skill-intro3}. A skill can package instructions, scripts, resources, and configuration files into a portable bundle that an agent can invoke when solving a class of tasks \cite{anthropic2025introducingskills,anthropic2026agentskillsdocs}.
However, a skill is rarely finished after its first release. As tools change, task distributions shift, and failures accumulate, skill authors revise instructions, patch scripts, add resources, etc. Skill evolution is part of how agent capabilities become reliable over time \cite{ExpeL,skill_evo1}.

Recent work on self-evolving agents has started to automate this process. 
A typical skill self-evolution loop executes an agent on tasks, collects trajectories, diagnoses failures, and revises the current skill accordingly~\cite{ExpeL,Skillopt}. 
This trajectory-driven paradigm grounds updates in task-specific execution evidence. 
However, its guidance is bounded by the quality and coverage of the trajectories collected in the current run~\cite{skillpro}. 
Noisy or low-quality trajectories may make it difficult to derive reliable guidance, while incomplete coverage of failure modes may cause the resulting updates to overfit to the limited observation. 
Consequently, trajectory-driven evolution may overlook broadly useful update knowledge that is not revealed by current executions~\cite{skill-intro4}.


A complementary source of evolution knowledge lies in public version histories. Software engineering research has mined historical patches to recover reusable edit and repair patterns~\cite{Getafix,FixMiner}.
Public repositories similarly preserve how developers revise instructions, scripts, and other components during maintenance. These changes may encode update strategies that are not observed in the current task run. 
Nevertheless, skill version histories have received limited attention as a source of reusable experience for skill self-evolution, and it remains unclear whether they provide knowledge beyond that already produced from trajectories \cite{skill-intro1,SkillsBench}.

To validate this premise, we conduct a motivation study that compares matched public and trajectory-derived skill changes, under a unified comparison protocol.
The comparison reveals complementary coverage: trajectory-derived changes are highly concentrated in instructions and a narrow set of intents and patterns, whereas public histories cover broader changes.
Meanwhile, trajectory-only changes remain, showing that public histories complement rather than replace trajectory-derived evidence.
This complementarity does not make historical changes directly usable. Raw diffs are fine-grained and repository-specific, and external experience may not always be relevant to the current failure.
Exploiting version histories therefore requires addressing two challenges: distilling raw changes into transferable evolution guidance and selectively fusing this guidance with the trajectory-derived proposal\cite{attention1,attention2}.




To address these challenges, we propose \textbf{\textit{\tool{}}}, a method using version-change experience (\textbf{\textit{VCE}}) to enhance \textbf{\textit{Skill}} self-evolution with two stages.
First, \emph{Experience Distillation} abstracts raw diffs into structured update events and progressively generalizes them into update patterns and insights, together forming an experience bank.
Second, \emph{Optimization with Adaptive Experience Attention} selects relevant experience from the bank while a compatible base evolver independently produces a self-proposal from the trajectories.
It then fuses the external experience and self-proposal with attention weights. After each optimization based on the fused proposal, adaptation feedback adjusts experience selection and attention allocation for the next iteration.

We evaluate \tool{} on 5 benchmarks with 4 LLMs by enhancing 3 skill self-evolution methods and comparing against 2 non-evolutionary baselines.
The experiments examine effectiveness, the ablation of \tool{}, and the cross-model transferability of evolved skills.
The results show that \tool{} consistently improves the base evolvers.

This paper makes the following contributions:
\begin{itemize}[leftmargin=*]
    \item We conduct the pilot study comparing public changes with trajectory-derived skill changes, revealing their complementary coverage and taking public VCE as an external prior for skill evolution.
    
    \item We introduce \tool{}, which distills raw version changes into reusable experience and adaptively selects and fuses it with trajectory-derived proposals during iterative skill optimization.

    \item We conduct extensive experiments, demonstrating the effectiveness and generality of \tool{}.
\end{itemize}

\section{Background and Related Work}
\label{sec:background}

\subsection{Agent Skills}

Following Anthropic's definition and the Agent Skills specification \cite{anthropic2025introducingskills,anthropic2026agentskillsdocs}, we view an agent skill as a reusable, filesystem-based capability package that an agent can discover and load on demand for a specialized task. At minimum, a skill is a directory containing a \texttt{SKILL.md} file.

We represent an agent skill as
\begin{equation}
    s = \langle m, \mathcal{C}_s \rangle,
    \qquad
    \mathcal{C}_s = \{c_k\}_{k=1}^{K_s},
    \label{eq:agent-skill}
\end{equation}

where $m$ denotes the metadata in the YAML frontmatter, including the required name and description, and $\mathcal{C}_s$ denotes the set of components contained in skill $s$. Each $c_k$ represents a functional component, such as procedural instructions, executable scripts, reference materials, configurations, templates, or other supporting artifacts. 
The instruction component corresponding to the body of \texttt{SKILL.md} is required, whereas supporting components are optional. During use, the agent first observes $m$ for skill discovery, loads the instruction component when the skill is activated, and accesses other task-relevant components in $\mathcal{C}_s$ as needed.

Prior work studies skill artifacts and libraries from complementary perspectives \cite{ContextMatters,skill-intro3,liu2026agentskillswildempirical}. 
SkillsBench \cite{SkillsBench} evaluates curated and self-generated skills. SkillReducer \cite{SkillReducer} compresses skill descriptions and bodies while preserving behavior. 
SKILLFOUNDRY \cite{SKILLFOUNDRY} constructs and iteratively validates skill libraries from heterogeneous scientific resources. 
Unlike work on evaluation, ecosystem analysis, compression, or resource-to-skill construction, we treat adjacent version changes themselves as reusable evolution experience.

\subsection{Skill Self-Evolution}

Skill self-evolution automates the improvement of a skill from execution feedback. A typical loop executes an agent on one or more tasks with the current skill, records the resulting trajectories, diagnoses observed behavior, proposes a candidate update, and evaluates the candidate on held-out tasks or regression cases before accepting it \cite{Reflexion,ExpeL}.

Let $s_t$ be the current skill and $B_t$ the task batch at evolution round $t$. Execution produces a trajectory set and an evolver generates a candidate skill:
\begin{equation}
    \mathcal{T}_t = \mathrm{Exec}(s_t, B_t), \qquad
    \tilde{s}_{t+1} = \mathrm{Evolve}(s_t, \mathcal{T}_t).
    \label{eq:skill-evolution}
\end{equation}

The candidate is evaluated using validation cases $V_t$, to determine the next skill:
\begin{equation}
    s_{t+1} = \mathrm{Select}(s_t, \tilde{s}_{t+1}; V_t).
    \label{eq:skill-selection}
\end{equation}

The operator selects $\tilde{s}_{t+1}$ only when the candidate satisfies the acceptance criteria on $V_t$; otherwise, it retains $s_t$. This recurrence captures the execution--diagnosis--update--validation loop.

Recent methods instantiate trajectory-driven skill self-evolution in different ways. EvoSkill \cite{EvoSkill} and SkillForge \cite{SkillForge} diagnose failed executions to revise skills, while SkillOpt \cite{Skillopt} and Skills-Coach \cite{Skills-Coach} optimize skill artifacts from scored rollouts or task-level evaluations. CoEvoSkills \cite{CoEvoSkills} couples multi-file skill generation with a verifier, SkillClaw \cite{SkillClaw} aggregates interactions across users, SkillOS \cite{SkillOS} learns a repository-curation policy from task streams, and EmbodiSkill \cite{EmbodiSkill} distinguishes skill defects from execution lapses in embodied trajectories. 
MetaSkill-Evolve \cite{MetaSkill-Evolve} further extends this paradigm by recursively improving both task skills and the meta-skills that govern their evolution through two execution-driven timescales.
However, their update signals arise mainly from current or accumulated task interactions and evaluations. 
In contrast, \tool{} mines public skill version histories as a previously untapped source of prior knowledge for skill evolution and integrates the distilled experience with trajectory-derived proposals.

\subsection{Evolution Knowledge from Version Histories}

Version histories have long supported reusable edit mining: studies of repetitive revisions, Getafix \cite{Getafix}, and FixMiner \cite{FixMiner} recover recurring code-change or repair patterns. However, these approaches model syntax-level code edits and repair contexts \cite{learn_exp,learn_exp2}, whereas agent skills are heterogeneous, multi-component artifacts whose changes encode semantic intents across instructions, scripts, references, and configurations.
Their code-specific representations are therefore not directly applicable to skill evolution. \tool{} extends history-based edit mining to agent skills by distilling guidance from cross-repository skill version changes and retrieving relevant experience to complement trajectory-derived proposals.

\section{Motivation Study}
\label{sec:motivation}

\begin{figure}[t]
    \centering
    \includegraphics[width=\linewidth]{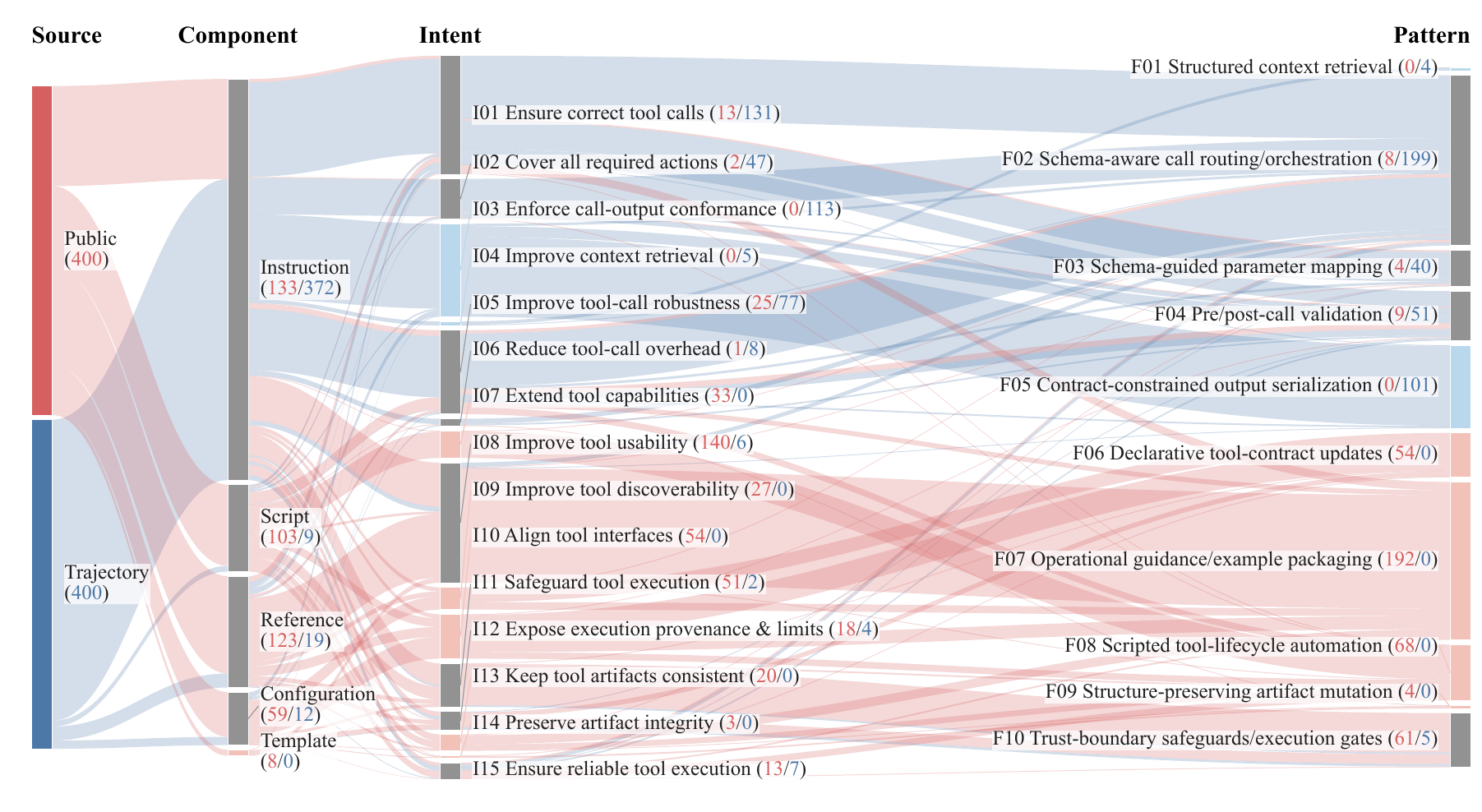}
    \vspace{-0.4em}
    {\small\textbf{(a)} BFCL v4 Full}
    \par\vspace{0.2em}
    \includegraphics[width=\linewidth]{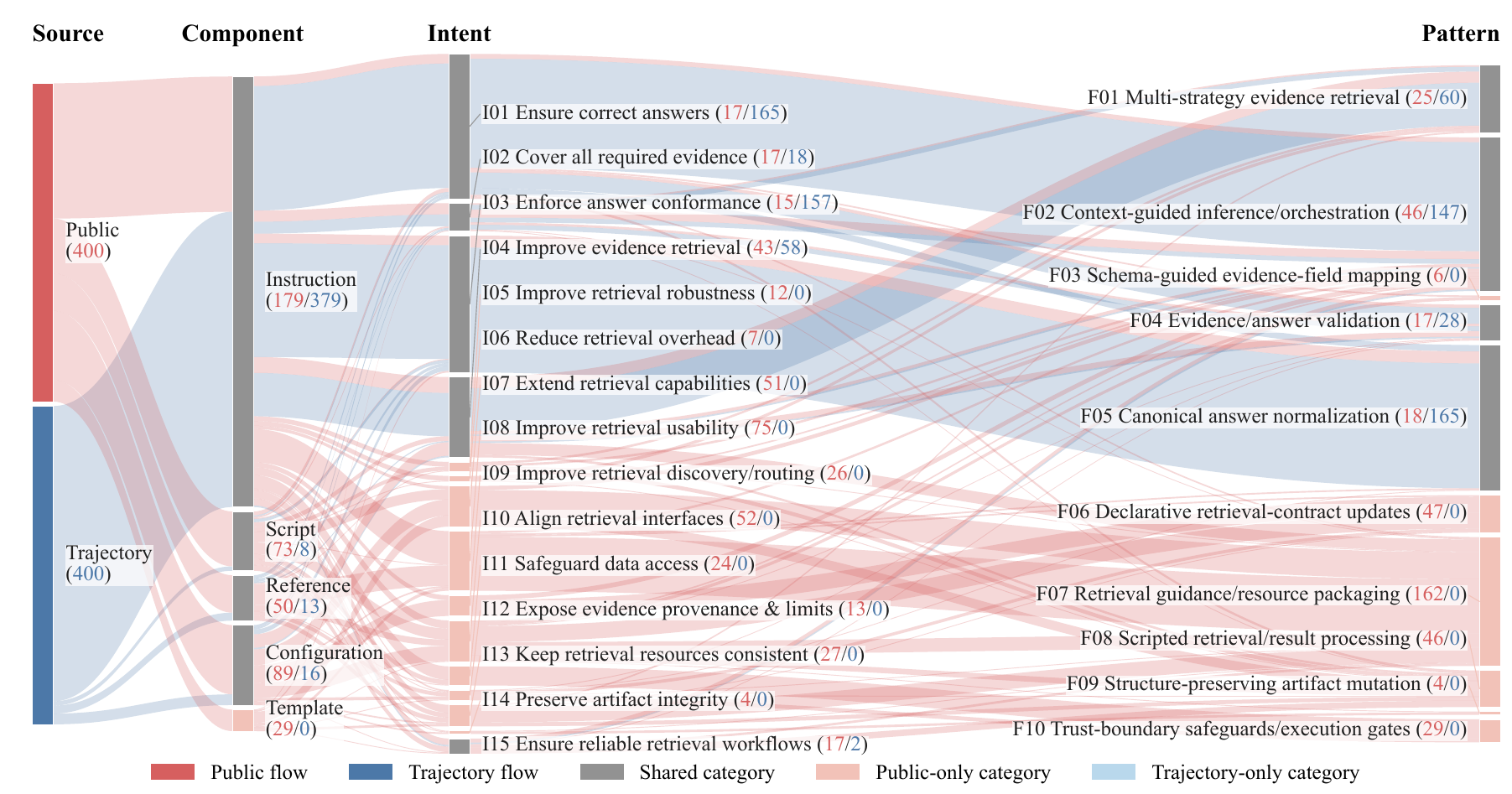}
    \vspace{-0.4em}
    {\small\textbf{(b)} SearchQA}
    \caption{Public and trajectory-derived skill changes.}
    \label{fig:motivation-vce-trajectory}
\end{figure}

Trajectory-based skill evolution derives update guidance from the trajectory collected during the optimization run.
However, the resulting guidance is sensitive to trajectory quality: noisy or incomplete trajectories can obscure failure causes and provide limited actionable guidance.
Moreover, this evidence is bounded by the failure modes exposed by the sampled rollouts. Consequently, useful update strategies that do not arise in the current executions may remain unobserved.
We provide a supplementary analysis of trajectory-quality sensitivity in the Appendix.

Public skill version histories provide potential complementary knowledge. They record skill changes adopted by developers during maintenance, and therefore constitute a practically grounded and rich source of evolution knowledge.
However, it remains unclear whether these public changes merely repeat the update strategies that trajectory-based evolution can produce adequately, or contain additional update knowledge.
This distinction directly determines the motivation for our method: if the two sources exhibit complementary update coverage, version histories can serve as an external prior for trajectory-based evolution.

We therefore ask: \emph{Do public skill version histories contain evolution knowledge that is not adequately captured by trajectory-based skill evolution?}
To answer this question, we compare the actual skill changes produced by the two evolution processes under a unified analysis protocol.

\subsection{Data and Comparison Protocol}

\paragraph{Data Sources.}
The \emph{public} source consists of changes between adjacent versions collected from GitHub and ClawHub.
The \emph{trajectory} source consists of changes produced through the evolution of several representative methods: EvoSkill~\cite{EvoSkill}, SkillClaw~\cite{SkillClaw}, and SkillOpt~\cite{Skillopt}.
For both sources, we analyze paired skills between changes.
We match the two sources within each task domain and construct a comparison corpus containing 400 skill-change units from each source.

\paragraph{Unified Comparison Protocol.}
We segment each skill pair into skill-change units by grouping edits that implement the same modification intent and pattern. Each unit may involve multiple components but is assigned one intent and one pattern.
An LLM judge annotates both sources using the same domain-specific taxonomy. The taxonomy distinguishes where an edit is made (component), why it is introduced (intent), and which reusable modification strategy it follows (pattern).
Multiple authors manually verify a sample of the annotations, with all reviewed cases passing the verification (100\%).
We compare the resulting distributions and identify categories observed in both sources as shared, and those observed in only one source as public-only or trajectory-only.
Full details are provided in the Appendix.

\subsection{Key Observations}

\paragraph{Complementary Coverage across Evolution Sources.}
Figure~\ref{fig:motivation-vce-trajectory} shows that trajectory-derived changes are highly concentrated: 372 of 400 BFCL units and 379 of 400 SearchQA units affect instructions. In contrast, public changes are distributed more broadly across different components.
At the semantic level, public changes also introduce intent and pattern absent from the trajectory source. BFCL contains 5 public-only intent categories and 4 public-only pattern families, while SearchQA contains 10 public-only intent categories and 6 public-only patterns.

These results show that public version histories do not merely duplicate the updates produced from online trajectories.
Though the two sources share a core set of update categories, public changes extend the scope of evolution beyond what is exposed by current rollouts.
Meanwhile, BFCL also contains trajectory-only intents and patterns, indicating that public version histories cannot replace the task-grounded evidence obtained during online evolution.

\paragraph{Design Implication.}
Overall, public version histories provide evolution knowledge that is not fully covered by trajectory-derived changes, while trajectory evidence remains directly grounded in the target task.
This complementarity motivates integrating public version-change experience as an external prior through adaptive fusion.

\section{\tool{} Method}
\label{sec:method}

\begin{figure*}[t]
    \centering
    \includegraphics[width=\textwidth]{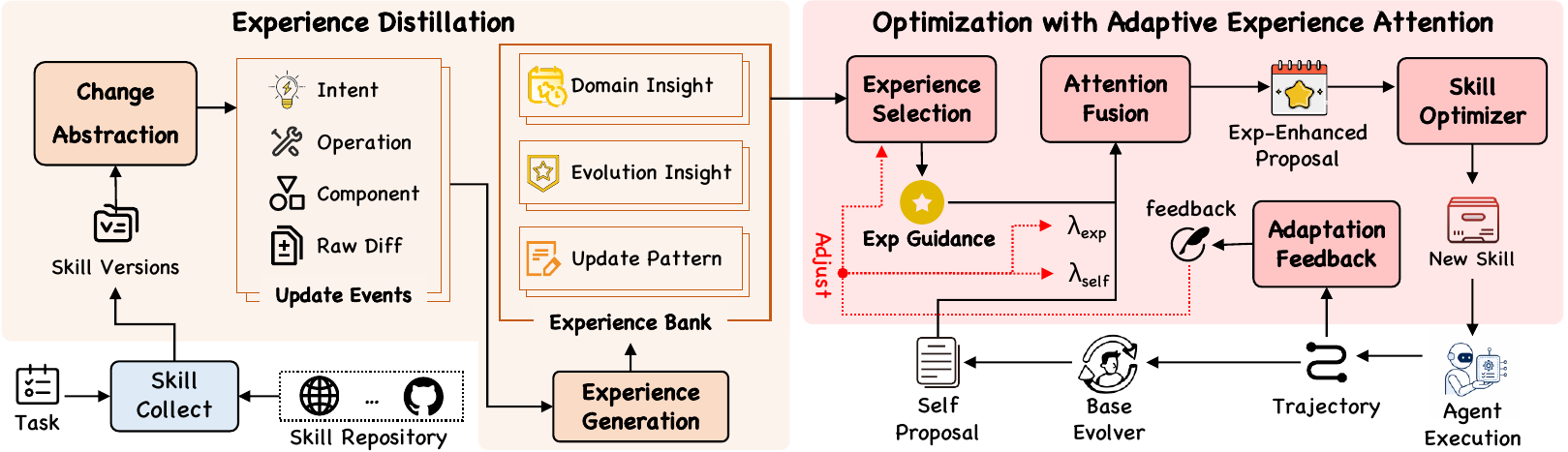}
    \caption{Overview of \tool{}. The left stage distills the version-change experience from public skill versions. The right stage enhances the skill self-evolution by allocating adaptive attention between external experience and trajectory-derived proposals.}
    \label{fig:method-overview}
\end{figure*}

Figure~\ref{fig:method-overview} presents \tool{}, a framework that transfers experience from historical skill versions to guide the self-evolution of a target skill. The framework consists of two stages. (1) \emph{Experience Distillation} distills historical version changes from public skill repositories into reusable experience and organizes it into a structured experience bank. (2) \emph{Optimization with Adaptive Experience Attention} selects task-relevant experience from this bank and adaptively combines it with the current skill's own evolution proposal to guide iterative skill improvement.
The details of the LLM judge used in \tool{} are provided in the Appendix.

\subsection{Experience Distillation}

Raw version diffs encode concrete, repository-specific implementations rather than directly reusable evolution guidance. We therefore abstract version changes into update events and distill them to constructive experience.
Figure~\ref{fig:experience-distill} provides a concrete example of how \tool{} converts a pair of skill versions into structured update events and organizes the distilled knowledge.

\subsubsection{Change Abstraction.}
As illustrated in Figure~\ref{fig:experience-distill}, file-level edits between two versions are converted into an update event that separates the concrete raw diff from its component, operation, and intent.
Specifically, for each collected skill, \tool{} orders its versions chronologically and compares each pair of adjacent versions to obtain a raw diff. It then decomposes each diff into individual changes, each of which is represented as an update event:
\begin{equation}
    u_j = \langle d_j, \{c_j\}, o_j, i_j \rangle,
    \label{eq:update-event}
\end{equation}
where $d_j$ is the \emph{raw diff segment} corresponding to the event, containing its specific editions. The field $\{c_j\}$ identifies the affected \emph{components}, such as instructions, scripts, etc. The field $o_j$ describes the semantic \emph{operation} performed by the change, such as adding an input-existence check. The field $i_j$ records the update \emph{intent}, such as preventing execution when a required input is missing.
This structured representation supports the subsequent abstraction of update events into reusable version-change experience.

\subsubsection{Experience Generation.}
\begin{figure}[th]
    \centering
    \includegraphics[width=\linewidth]{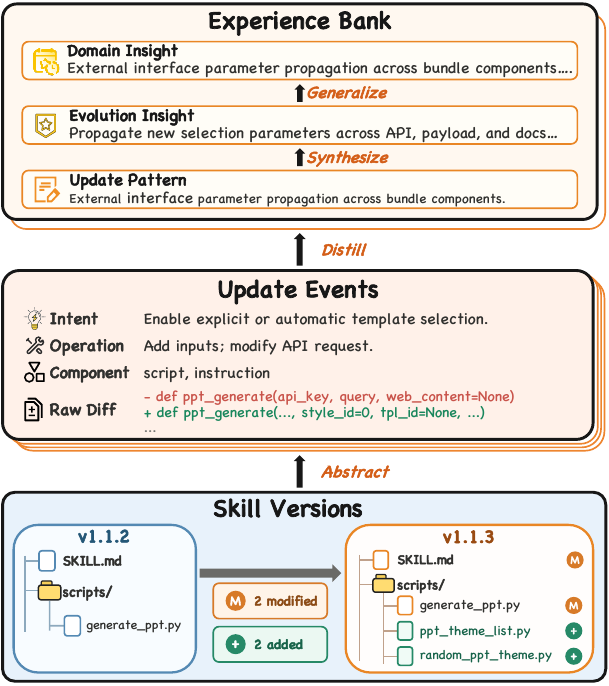}
    \caption{
    Illustration of experience distillation.
    }
    \label{fig:experience-distill}
\end{figure}

As shown in Figure~\ref{fig:experience-distill}, 
\tool{} then transforms update events into reusable knowledge at three levels.
First, within each skill's version history, it groups related events based on their components, operations, and intents, removes repository-specific details, and summarizes each group as an \emph{update pattern}.
Second, it synthesizes the resulting patterns for each skill into \emph{evolution insights} that characterize recurring update strategies in that skill's history.
Third, it generalizes common evolution insights across skills in the same task domain into \emph{domain insights} that capture transferable principles for skill optimization.

The resulting entries from all collected skills are aggregated into the experience bank $\mathcal{B}$. It contains multiple entries at each level, spanning different skills and task domains. During skill self-optimization, the bank serves as an external knowledge corpus from which \tool{} selects experiences relevant to the current task domain and skill state.
All LLM-based distillation steps use fixed prompt and structured outputs; the details are provided in supplementary materials.

\subsection{Optimization with Adaptive Experience Attention}

\subsubsection{Experience Selection.}
At iteration $t > 1$, \tool{} constructs a selection context from the target task, the current skill, and the unified feedback from the preceding update:
\begin{equation}
    z_t=\mathrm{Context}(\mathcal{T},S_{t-1},f_{t-1}).
    \label{eq:selection-context}
\end{equation}
where $\mathcal{T}$ denotes the target task, $S_{t-1}$ is the current skill, and $f_{t-1}$ is the feedback from the preceding update, which will be introduced later.
The $\mathrm{Context}(\cdot)$ function organizes them into a structured textual representation $z_t$ that serves as the query for experience selection.

\tool{} then uses an LLM-based selector to semantically compare $z_t$ with the entries in $\mathcal{B}$ and select at most $K$ experiences:
\begin{equation}
    \mathcal{X}_t
    =\mathrm{Select}_{\leq K}(\mathcal{B},z_t),
    \qquad
    p^{\mathrm{exp}}_t=\mathrm{Aggregate}(\mathcal{X}_t),
    \label{eq:experience-selection}
\end{equation}
where $K$ is the selection budget and $p^{\mathrm{exp}}_t$ is the compact external experience guidance aggregated from the selected entries. 
Selection remains conditioned primarily on compatibility with the current task and skill, while $f_{t-1}$ adjusts the selector's preference for external experience. Positive feedback favors external guidance, whereas negative feedback makes external selection more conservative. This makes selection adaptive across iterations: after the skill is updated, the new skill state and unified feedback change $z_{t+1}$ and therefore the selected experiences.


\subsubsection{Attention Fusion for Skill Optimizer.}
In parallel, the base evolver uses the current skill and its execution trajectory to produce a self proposal:
\begin{equation}
    p^{\mathrm{self}}_t
    =\mathrm{BaseEvolver}(S_{t-1},\tau_{t-1}),
    \label{eq:self-proposal}
\end{equation}
where $\tau_{t-1}$ is the trajectory produced by executing $S_{t-1}$. The base evolver identifies potential improvements from the successes, failures, and optimization needs reflected in this trajectory, and proposes edits for producing $S_t$. Thus, the self proposal is grounded in the current skill's own execution, whereas $p^{\mathrm{exp}}_t$ provides reusable guidance distilled from external skill histories.

Adaptive attention assigns source-level weights to external experience guidance and the self proposal:
\begin{equation}
    \lambda^{(t)}_{\mathrm{exp}}
    +\lambda^{(t)}_{\mathrm{self}}=1,
    \qquad
    0\leq
    \lambda^{(t)}_{\mathrm{exp}},
    \lambda^{(t)}_{\mathrm{self}}
    \leq 1.
    \label{eq:attention-simplex}
\end{equation}

The weights serve as prompt-level reliance instructions for the LLM-based fuser, rather than as probabilities.
Both weights are initialized to \(0.5\). The fuser combines compatible guidance and prioritizes the higher-weight source when the two sources conflict.
It produces:
\begin{equation}
    p^{\mathrm{enh}}_t
    =\mathrm{Fuse}\!\left(
        p^{\mathrm{exp}}_t,p^{\mathrm{self}}_t;
        \lambda^{(t)}_{\mathrm{exp}},
        \lambda^{(t)}_{\mathrm{self}}
    \right),
    \label{eq:enhanced-proposal}
\end{equation}
where $p^{\mathrm{enh}}_t$ is the experience-enhanced proposal, represented as a structured set of candidate edits. 

To preserve source evidence for subsequent feedback, each selected external experience and each self-proposal item is assigned a unique provenance identifier, such as \texttt{EXP-k} or \texttt{SELF-k}, which will be preserved in the fusion.
These edit-level provenance tags are later compared with the actual skill update to determine which proposal items were realized and whether the implemented update was primarily supported by external experience or by the self proposal.

The skill optimizer applies the enhanced proposal to the current skill:
\begin{equation}
    S_{t}=\mathbf{O}(S_{t-1},p^{\mathrm{enh}}_t).
    \label{eq:skill-optimizer}
\end{equation}

Depending on the proposal, the LLM-based optimizer revises the skill bundle to produce $S_t$.

\subsubsection{Adaptation Feedback.}
The agent then executes $S_t$, producing a new trajectory $\tau_t$ for the base evolver in the next iteration. Separately, the completed skill update and its validation produce two intermediate signals: source feedback $f^S_t$, which summarizes whether the implemented edits rely more on external experience or on the self proposal, and validation feedback $f^V_t$, which measures whether the updated skill improves validation performance.

Because the optimizer may realize only part of the enhanced proposal, \tool{} compares the actual skill change with the proposal's provenance-tagged edits:
\begin{equation}
    \Delta S_t=\mathrm{Diff}(S_t,S_{t-1}),
    \label{eq:source-attribution}
\end{equation}

Let $\mathcal{C}_t=\{c_{t,k}\}_{k=1}^{n_t}$ denote the provenance-tagged candidate edits in $p_t^{\mathrm{enh}}$. The attribution function semantically aligns each candidate edit with the realized changes and assigns an item-level attribution:
\begin{equation}
    a_{t,k}
    =\mathrm{Attr}(\Delta S_t,c_{t,k}),
    \label{eq:item-attribution}
\end{equation}
where $a_{t,k}\in\{-1,0,+1\}$ for $k=1,\ldots,n_t$. We set $a_{t,k}=+1$ when an externally supported candidate edit is realized, $a_{t,k}=-1$ when a self-supported candidate edit is realized, and $a_{t,k}=0$ when the candidate edit is not realized, cannot be reliably matched, or has mixed provenance. The source feedback is the mean attribution over all candidate edits:
\begin{equation}
    f^S_t=\frac{1}{n_t}\sum_{k=1}^{n_t}a_{t,k},
    \qquad f^S_t\in[-1,1].
    \label{eq:source-feedback}
\end{equation}

Positive values indicate greater realized contribution from external experience, negative values indicate greater contribution from the self proposal, and values near zero indicate balanced, weak, or uncertain attribution.

Second, validation feedback $f^V_t$ represents the change in skill performance on a fixed validation set $V$:
\begin{equation}
f^V_t=
\left\{
\begin{array}{ll}
    +1, & \Delta^V_t > 0,\\
    -1, & \Delta^V_t \leq 0,
\end{array}
\right. 
\label{eq:validation-feedback}
\end{equation}
where $\Delta^V_t$ denotes the performance change of validation.
We combine source attribution and validation performance into a unified source-preference feedback:
\begin{equation}
    f_t=f^V_t f^S_t,
    \qquad f_t\in[-1,1].
    \label{eq:unified-feedback}
\end{equation}
Positive $f_t$ shifts the next iteration toward external experience, negative $f_t$ shifts it toward the self proposal, and values near zero leave the preference nearly unchanged. The same feedback is included in the next selection context and used to update the attention weights:
\begin{equation}
\begin{array}{rcl}
    \lambda^{(t+1)}_{\mathrm{exp}}
    &=&\mathrm{clip}\!\left(
        \lambda^{(t)}_{\mathrm{exp}}+\eta f_t,
        0.3, 0.7
    \right),\\[2pt]
    \lambda^{(t+1)}_{\mathrm{self}}
    &=&1-\lambda^{(t+1)}_{\mathrm{exp}},
\end{array}
    \label{eq:attention-update}
\end{equation}
where the adjustment step size is fixed to $\eta=0.1$, and clipping to $[0.3,0.7]$ prevents either source from being permanently ignored. When an improving update relies more on one source, the method increases reliance on that source in proportion to $|f^S_t|$. When a non-improving update relies more on one source, it shifts reliance toward the other.

The loop is modular with respect to the base evolver. Different trajectory-driven self-evolution methods can supply $q^{\mathrm{self}}_t$ without changing the experience bank, feedback-conditioned selector, or source-level attention update.

\begin{table*}[ht]
\centering
\scriptsize
\renewcommand{\arraystretch}{0.8}
\newcommand{\rqgain}[1]{\textcolor{red}{(+#1)}}
\newcommand{\rqloss}[1]{\textcolor{green!50!black}{(-#1)}}
\resizebox{\textwidth}{!}{
\begin{tabular}{l | l | l l l l l l}
\toprule
\multicolumn{1}{c|}{Model} & \multicolumn{1}{c|}{Skill} & SearchQA & OfficeQA & ALFWorld & Spreadsheet & BFCL-v4 & Avg. \\
\midrule
& No Skill & 68.42 & 33.79 & 52.87 & 26.93 & 54.75 & 47.35 \\
& LLM Skill & 67.29 & 33.41 & 54.74 & 29.48 & 55.12 & 48.01 \\
& EvoSkill & 67.08 & 34.13 & 57.46 & 39.45 & 58.45 & 51.31 \\
\rowcolor{gray!15}
\cellcolor{white} & \textit{EvoSkill+VCE} & 70.46\,\rqgain{3.38} & 41.79\,\rqgain{7.66} & 59.13\,\rqgain{1.67} & 45.13\,\rqgain{5.68} & 60.71\,\rqgain{2.26} & 55.44\,\rqgain{4.13} \\
& SkillClaw & 69.41 & 36.12 & 59.01 & 42.46 & 61.23 & 53.65 \\
\rowcolor{gray!15}
\cellcolor{white} & \textit{SkillClaw+VCE} & 72.48\,\rqgain{3.07} & 41.48\,\rqgain{5.36} & 62.48\,\rqgain{3.47} & 46.73\,\rqgain{4.27} & \textbf{63.77}\,\rqgain{2.54} & 57.39\,\rqgain{3.74} \\
& SkillOpt & 74.27 & 46.12 & 59.70 & 49.08 & 54.82 & 56.80 \\
\rowcolor{gray!15}
\cellcolor{white}\multirow{-8}{*}{Qwen3.5-27B} & \textit{SkillOpt+VCE} & \textbf{78.86}\,\rqgain{4.59} & \textbf{49.42}\,\rqgain{3.30} & \textbf{63.43}\,\rqgain{3.73} & \textbf{52.14}\,\rqgain{3.06} & 57.52\,\rqgain{2.70} & \textbf{60.27}\,\rqgain{3.47} \\
\midrule
& No Skill & 72.67 & 36.57 & 69.76 & 41.07 & 51.13 & 54.24 \\
& LLM Skill & 73.75 & 43.49 & 71.04 & 41.84 & 53.44 & 56.71 \\
& EvoSkill & 76.16 & 41.28 & 70.45 & 43.96 & 59.11 & 58.19 \\
\rowcolor{gray!15}
\cellcolor{white} & \textit{EvoSkill+VCE} & 80.71\,\rqgain{4.55} & 48.41\,\rqgain{7.13} & 74.84\,\rqgain{4.39} & 47.02\,\rqgain{3.06} & 64.85\,\rqgain{5.74} & 63.17\,\rqgain{4.98} \\
& SkillClaw & 79.40 & 52.09 & 76.94 & 51.06 & 60.42 & 63.98 \\
\rowcolor{gray!15}
\cellcolor{white} & \textit{SkillClaw+VCE} & 84.14\,\rqgain{4.74} & 55.49\,\rqgain{3.40} & 80.40\,\rqgain{3.46} & 56.41\,\rqgain{5.35} & 64.17\,\rqgain{3.75} & 68.12\,\rqgain{4.14} \\
& SkillOpt & 83.11 & 52.16 & 81.06 & 54.79 & 64.16 & 67.06 \\
\rowcolor{gray!15}
\cellcolor{white}\multirow{-8}{*}{GPT-5.2} & \textit{SkillOpt+VCE} & \textbf{85.86}\,\rqgain{2.75} & \textbf{56.71}\,\rqgain{4.55} & \textbf{84.58}\,\rqgain{3.52} & \textbf{59.60}\,\rqgain{4.81} & \textbf{67.47}\,\rqgain{3.31} & \textbf{70.84}\,\rqgain{3.78} \\
\midrule
& No Skill & 70.46 & 35.54 & 66.24 & 29.46 & 53.41 & 51.02 \\
& LLM Skill & 70.20 & 34.16 & 69.67 & 30.79 & 55.37 & 52.04 \\
& EvoSkill & 72.04 & 38.16 & 68.06 & 32.72 & 56.15 & 53.43 \\
\rowcolor{gray!15}
\cellcolor{white} & \textit{EvoSkill+VCE} & 74.97\,\rqgain{2.93} & 42.53\,\rqgain{4.37} & 71.64\,\rqgain{3.58} & 35.48\,\rqgain{2.76} & 58.61\,\rqgain{2.46} & 56.65\,\rqgain{3.22} \\
& SkillClaw & 73.47 & 41.06 & 69.40 & 34.94 & 59.27 & 55.63 \\
\rowcolor{gray!15}
\cellcolor{white} & \textit{SkillClaw+VCE} & 76.98\,\rqgain{3.51} & 42.68\,\rqgain{1.62} & 72.87\,\rqgain{3.47} & 37.48\,\rqgain{2.54} & 64.16\,\rqgain{4.89} & 58.83\,\rqgain{3.20} \\
& SkillOpt & 73.82 & 45.55 & 68.08 & 37.78 & 61.36 & 57.32 \\
\rowcolor{gray!15}
\cellcolor{white}\multirow{-8}{*}{DeepSeek-v3.2} & \textit{SkillOpt+VCE} & \textbf{77.68}\,\rqgain{3.86} & \textbf{49.01}\,\rqgain{3.46} & \textbf{73.70}\,\rqgain{5.62} & \textbf{42.02}\,\rqgain{4.24} & \textbf{64.58}\,\rqgain{3.22} & \textbf{61.40}\,\rqgain{4.08} \\
\midrule
& No Skill & 76.40 & 51.56 & 73.16 & 45.06 & 57.42 & 60.72 \\
& LLM Skill & 76.46 & 50.77 & 74.14 & 50.05 & 58.76 & 62.04 \\
& EvoSkill & 74.83 & 52.03 & 72.45 & 55.16 & 59.57 & 62.81 \\
\rowcolor{gray!15}
\cellcolor{white} & \textit{EvoSkill+VCE} & 78.92\,\rqgain{4.09} & 56.22\,\rqgain{4.19} & 77.94\,\rqgain{5.49} & 59.70\,\rqgain{4.54} & 63.98\,\rqgain{4.41} & 67.35\,\rqgain{4.54} \\
& SkillClaw & 81.02 & 53.21 & 78.10 & 50.43 & 65.93 & 65.74 \\
\rowcolor{gray!15}
\cellcolor{white} & \textit{SkillClaw+VCE} & 84.56\,\rqgain{3.54} & 56.88\,\rqgain{3.67} & 82.04\,\rqgain{3.94} & 55.30\,\rqgain{4.87} & 68.37\,\rqgain{2.44} & 69.43\,\rqgain{3.69} \\
& SkillOpt & 82.71 & 54.17 & 83.12 & 57.91 & 66.06 & 68.79 \\
\rowcolor{gray!15}
\cellcolor{white}\multirow{-8}{*}{Claude Sonnet 5} & \textit{SkillOpt+VCE} & \textbf{87.13}\,\rqgain{4.42} & \textbf{58.44}\,\rqgain{4.27} & \textbf{85.70}\,\rqgain{2.58} & \textbf{62.04}\,\rqgain{4.13} & \textbf{70.04}\,\rqgain{3.98} & \textbf{72.67}\,\rqgain{3.88} \\
\bottomrule
\end{tabular}
}
\caption{Performance comparison on different models and benchmarks. For VCE-enhanced variants, red (green) parenthesized values indicate improvements (declines) over their corresponding baseline. Bold indicates the best result.}
\label{tab:rq1}
\end{table*}

\begin{figure}[t]
    \centering
    \includegraphics[width=\linewidth]{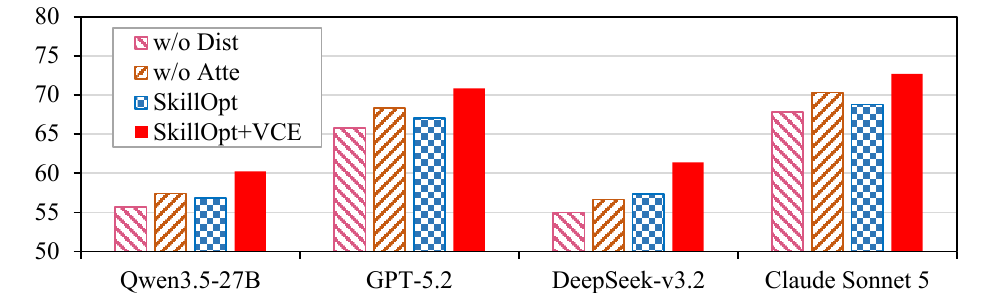}
    \caption{The performance of different variants of \tool{}.}
    \label{fig:rq2-ablation}
\end{figure}

\section{Experimental Evaluation}
\label{sec:evaluation}

\subsection{Experimental Setup}

\paragraph{Benchmarks and Models.}
We evaluate on five benchmarks that cover complementary forms of agent skill use: SearchQA~\cite{SearchQA} for open-domain question answering under noisy retrieval; OfficeQA~\cite{OfficeQA} for document, table, and numerical reasoning; ALFWorld~\cite{ALFWorld} for long-horizon embodied task execution; Spreadsheet~\cite{SpreadsheetBench} for spreadsheet operations and calculation; and BFCL-v4~\cite{BFCL} for function calling and tool selection. 

We take four models as agent: Qwen3.5-27B \cite{qwen35}, GPT-5.2 \cite{gpt52}, DeepSeek-v3.2 \cite{DeepSeek-V3.2}, and claude-sonnet-5 \cite{sonnet5}.

\paragraph{Compared baselines.}
We compare \tool{} against five groups of baselines. No Skill executes the target agent without loading any skill. LLM Skill uses a one-shot skill generated by an LLM without iterative refinement. 
EvoSkill \cite{EvoSkill} evolves skills from failure trajectories. 
SkillClaw \cite{SkillClaw} represents collective or multi-skill evolution. 
SkillOpt \cite{Skillopt} treats the skill as an optimizable textual state and accepts bounded edits through a validation gate.
\tool{} enhances these iterative base self-evolvers to compare.

\paragraph{Implementation and Evaluation Details.}
Each experimental configuration is independently run three times, and we report the mean results in the main paper.
We conduct sensitivity analyses of the key hyperparameters.
To assess reliability, we manually audit all LLM-based judging.
The whole results and details, such as the prompts, LLMs, and environments, are provided in the Appendix.

\subsection{Performance of \tool{}.}

Table~\ref{tab:rq1} presents paired comparisons between each self-evolution method and its VCE-enhanced variant. The gray rows denote variants augmented with \tool{}, while the parenthesized values show their improvements or declines relative to the corresponding non-VCE baseline. 
Introducing VCE improves every task score of each base evolver. It also increases the average score by 3.20--4.98 points. These results show that the benefit of \tool{} is consistent across different skill-evolution frameworks rather than being specific to a particular method.

These improvements arise because \tool{} leverages external evolution experience that is complementary to the agent's own trajectory-derived knowledge. Experience distillation makes this knowledge reusable and actionable, while adaptive experience selection and attention allocation improve its alignment with the current evolution process.

For each benchmark, \tool{} incurs approximately 0.5M tokens for the one-time experience distillation and 0.6M tokens for a complete evolution comprising approximately 20 iterations, resulting in a total additional cost of approximately 1.1M tokens. This corresponds to approximately 10\% additional token overhead relative to the best-performing baseline.

\subsection{Ablation Study.}
\paragraph{Variants.}
We conduct the ablation study and construct two variants of \tool{}, each removing one key component.
\textit{\textbf{w/o Dist}} removes experience distillation and directly uses raw version-change records as external evolution knowledge.
\textit{\textbf{w/o Atte}} replaces adaptive experience attention with a fixed equal-weight combination of the distilled external experience and the self proposal generated by base evolver.

\paragraph{Results.}
Figure \ref{fig:rq2-ablation} reports the average score on all benchmarks of removing different components of {\tool}, using the best baseline, SkillOpt, as the base evolver.
Removing experience distillation causes the largest performance degradation, with \textit{\textbf{w/o Dist}} even falling below SkillOpt on all four models. Raw version-change records contain implementation details that may not transfer directly to the current skill.
Experience distillation abstracts these records into reusable patterns and insights, thereby reducing irrelevant information and providing more actionable guidance.
Replacing adaptive attention with fixed equal weights also consistently underperforms {\tool}, although \textit{\textbf{w/o Atte}} still outperforms SkillOpt on three of the four models. This result indicates that distilled external experience is generally useful, but its contribution should not remain constant throughout evolution. As the skill state changes across iterations, the relevance of external experience and the reliability of the self proposal also vary. Adaptive experience attention accounts for these and allocates the two types of knowledge accordingly, enabling more targeted and reliable skill updates.

\subsection{Transferability of Skills.}

This experiment investigates whether skills produced through self-evolution remain effective when transferred across models. We evolve two skills on the same source model using SkillOpt and SkillOpt augmented with \tool{}, respectively, and then transfer them to other models for evaluation. Table~\ref{tab:skill-transferability} reports their transferred performance when Qwen3.5-27B is used as the source model.

\begin{table}[ht]
\centering
\scriptsize
\newcommand{\xfergain}[1]{\textcolor{red}{(+#1)}}
\setlength{\tabcolsep}{2pt}
\resizebox{\linewidth}{!}{
\begin{tabular}{l | l | l l l l l l}
\toprule
\multicolumn{1}{l|}{Target Model} & \multicolumn{1}{l|}{Source Skill} & SearchQA & OfficeQA & ALFWorld & Spreadsheet & BFCL-v4 & Avg. \\
\midrule
& SkillOpt & 78.01 & 47.47 & 79.16 & 52.11 & 64.10 & 64.17 \\
\rowcolor{gray!15}
\cellcolor{white}\multirow{-2}{*}{GPT-5.2} & \textit{SkillOpt+VCE} & \textbf{83.24} & \textbf{55.34} & \textbf{86.05} & \textbf{57.89} & \textbf{68.54} & \textbf{70.21}\,\xfergain{6.04} \\
\midrule
& SkillOpt & 71.16 & 44.94 & 69.60 & 38.16 & 62.10 & 57.19 \\
\rowcolor{gray!15}
\cellcolor{white}\multirow{-2}{*}{DeepSeek-v3.2} & \textit{SkillOpt+VCE} & \textbf{78.12} & \textbf{49.53} & \textbf{75.11} & \textbf{43.94} & \textbf{65.48} & \textbf{62.44}\,\xfergain{5.25} \\
\midrule
& SkillOpt & 81.10 & 53.03 & 83.69 & 56.01 & 66.42 & 68.05 \\
\rowcolor{gray!15}
\cellcolor{white}\multirow{-2}{*}{Claude Sonnet 5} & \textit{SkillOpt+VCE} & \textbf{87.37} & \textbf{58.14} & \textbf{86.80} & \textbf{61.73} & \textbf{71.97} & \textbf{73.20}\,\xfergain{5.15} \\
\bottomrule
\end{tabular}
}
\caption{Cross-model transfer performance of skills.}
\label{tab:skill-transferability}
\end{table}

\paragraph{Results.}
The skill evolved with \tool{} outperforms its SkillOpt counterpart. Specifically, \tool{} improves the average scores of the three target models by 6.04, 5.25, and 5.15 points, respectively. 
Notably, this transfer gain exceeds the gain over the available paired non-VCE counterparts in Table~\ref{tab:rq1}. 
It demonstrates that the skills evolved by \tool{} possess stronger cross-model transferability.

We attribute it to the more generalizable evolution knowledge incorporated by \tool{}. Trajectory-driven evolution primarily learns from source-model trajectories and may therefore retain model-specific behavior patterns. 
In contrast, \tool{} introduces reusable external experience. By suppressing model-specific trajectory details and incorporating generalizable task-solving principles, this external experience makes the evolved skill less dependent on the source model and more reliable when applied to other models.


\section{Conclusion}
\label{sec:conclusion}
In this work, we propose \tool{}, which enhances skill self-evolution with public skill versions.  
Our pilot study shows that public changes and those derived from trajectory-derived self-evolution provide complementary coverage.  
\tool{} distills public skill changes into structured prior experience and adaptively fuses it with trajectory-derived proposals from the base evolver, combining reusable prior knowledge with task-specific execution evidence to guide skill evolution.  
Experiments across five benchmarks, four LLMs, and three base evolvers demonstrate that \tool{} effectively improves skill self-evolution and yields skills with stronger cross-model transferability in the evaluated setting.  
Overall, our work highlights public skill version histories as an underexplored yet effective source of prior knowledge and provides a practical experience-augmented extension to the trajectory-driven paradigm of skill self-evolution.

\bibliography{aaai2027}



\bigskip

\clearpage
\setcounter{page}{1}
\appendix
\setcounter{figure}{0}
\setcounter{table}{0}
\setcounter{algorithm}{0}

\definecolor{appendixdiffred}{HTML}{B2182B}
\definecolor{appendixdiffgreen}{HTML}{1B7837}
\definecolor{appendixpromptgray}{HTML}{ECEDEF}
\definecolor{appendixrulegray}{HTML}{A7ABB0}
\lstdefinestyle{appendixprompt}{
  basicstyle=\ttfamily\fontsize{7.25}{8.8}\selectfont,
  backgroundcolor=\color{appendixpromptgray},
  frame=single,
  framerule=0pt,
  framesep=7pt,
  rulecolor=\color{appendixpromptgray},
  xleftmargin=7pt,
  xrightmargin=7pt,
  aboveskip=0pt,
  belowskip=0pt,
  breaklines=true,
  breakatwhitespace=true,
  columns=fullflexible,
  keepspaces=true,
  showstringspaces=false,
  upquote=true,
  numbers=none
}
\newcommand{\appendixdiffbox}[1]{\colorbox{appendixpromptgray}{\parbox{\dimexpr\linewidth-2\fboxsep\relax}{#1}}}

\twocolumn[
\begin{center}
    {\LARGE\bfseries Appendix for \tool{}}
    \vspace{0.4em}
\end{center}
]
\raggedbottom
\makeatletter
\setlength{\@dblfptop}{0pt}
\setlength{\@dblfpsep}{12pt plus 2fil}
\setlength{\@dblfpbot}{0pt plus 1fil}
\makeatother
\suppressfloats[t]

\noindent
This supplementary appendix provides the motivation-study protocol, implementation details, reproducibility information, extended analyses, and structured interfaces referenced by the main paper.
It is organized around three aspects: the motivation study, the mechanism of \tool{}, and the experimental details.

\section{Motivation Study Protocol and Supplementary Analyses}
\label{app:motivation}

\subsection{Matched Source Comparison}

The matched study examines how public version histories and trajectory-driven evolution differ in the skill-update knowledge they expose.
Both sources use the same skill-change unit and the same component, intent, and pattern taxonomy.

\paragraph{Public source.}
For BFCL and SearchQA, we collect skills from GitHub and ClawHub whose descriptions are semantically relevant to the benchmark and whose histories contain at least ten version updates.
We remove adjacent versions with no substantive skill change.
The retained public pool contains 38 skills and 1,266 updates for BFCL, and 21 skills and 1,089 updates for SearchQA.

\paragraph{Trajectory source.}
We run EvoSkill~\cite{EvoSkill}, SkillClaw~\cite{SkillClaw}, and SkillOpt~\cite{Skillopt} on BFCL and SearchQA.
Each method contributes 400 iterative updates across the two benchmarks, producing 1,200 trajectory-derived updates in total.
These updates follow the execution evidence observed by the current agent during skill evolution.

\paragraph{Balanced comparison.}
From the two source pools, we construct a benchmark-matched corpus containing 400 skill-change units from each source.
The shared protocol supports direct comparison of component, intent, and pattern distributions.

\subsection{Skill-Change Units and Annotation Taxonomy}
\label{app:change-unit}
\label{app:motivation-taxonomy}

Let $(s^{-},s^{+})$ denote the skill before and after one public or trajectory-derived update, and let $D(s^{-},s^{+})$ be their file-level difference.
A \emph{skill-change unit} is the maximal set of edits that implements one semantic intent through one reusable modification pattern.
Functionally coupled edits may span several files or components, while edits with distinct intents form separate units.
Formatting-only changes, generated files, and dependency-lock churn are removed before segmentation.

Each unit is described by three dimensions.
The \emph{component} records the affected artifact types, the \emph{intent} records the objective of the update, and the \emph{pattern} records the reusable modification strategy.
Components are multi-label, whereas intent and pattern each receive one primary label.

\begin{table*}[t]
\centering
\small
\begin{tabular}{p{0.18\textwidth}p{0.27\textwidth}p{0.45\textwidth}}
\toprule
Dimension & Annotation target & Decision rule \\
\midrule
Component & Affected skill artifacts & Assign all affected artifact types, including instructions, scripts, references, configurations, and templates. \\
Intent & Objective of the update & Assign the primary outcome pursued by the change. \\
Pattern & Modification strategy & Abstract the operational mechanism into a repository-independent family. \\
\bottomrule
\end{tabular}
\vspace{0.65em}

\noindent\textit{Selected intent labels and pattern families used in the representative cases.}

\begin{tabular}{p{0.16\textwidth}p{0.28\textwidth}p{0.46\textwidth}}
\toprule
Type & Label & Definition \\
\midrule
Intent & \texttt{INTENT\_04}: Evidence retrieval quality & Improve the precision, recall, navigation, or grounding of evidence acquisition before reasoning or execution. \\
Intent & \texttt{INTENT\_10}: Compatibility and interoperability & Align the skill with supported platforms, dependencies, schemas, clients, or execution environments. \\
Pattern & \texttt{FAMILY\_01}: Evidence retrieval and navigation & Modify how evidence is searched, located, traversed, or extracted from information sources. \\
Pattern & \texttt{FAMILY\_08}: Executable workflow automation & Modify executable helpers that automate requests, lifecycle steps, result enrichment, or artifact generation. \\
\bottomrule
\end{tabular}
\caption{Annotation dimensions and selected taxonomy labels used in the representative skill-change units.}
\label{tab:taxonomy-dimensions}
\label{tab:selected-taxonomy}
\end{table*}

\subsection{Representative Change Units}

Figure~\ref{fig:representative-change-units} contrasts a coordinated public-history update with a trajectory-derived instruction update.
The public update changes both user-facing guidance and executable code to maintain service compatibility.
The trajectory-derived update adds extraction rules for a failure format observed during SearchQA evolution.

\begin{figure*}[t]
\centering
\setlength{\fboxsep}{7pt}
\setlength{\tabcolsep}{3pt}
\renewcommand{\arraystretch}{1.12}
\begin{minipage}[t]{0.485\textwidth}
\raggedright
\textbf{(a) Public-history change}\hfill{\small BFCL $\cdot$ \texttt{plan2meal}}\\[-0.25em]
{\color{appendixrulegray}\rule{\linewidth}{0.5pt}}
\vspace{0.45em}

\appendixdiffbox{%
{\bfseries\footnotesize SKILL.md}\\[0.25em]
{\ttfamily\fontsize{7.2}{8.8}\selectfont
\textcolor{appendixdiffred}{- Default backend: https://\textless BACKEND\textgreater.convex.site}\\
\textcolor{appendixdiffgreen}{+ Default backend: https://\textless BACKEND\textgreater.convex.cloud}\\
\textcolor{appendixrulegray}{...}}

\vspace{0.65em}
{\bfseries\footnotesize index.ts}\\[0.25em]
{\ttfamily\fontsize{7.2}{8.8}\selectfont
\textcolor{appendixdiffred}{- convexUrl: process.env.CONVEX\_URL}\\
\textcolor{appendixdiffred}{- \ \ || 'https://\textless BACKEND\textgreater.convex.site',}\\
\textcolor{appendixdiffgreen}{+ convexUrl: process.env.CONVEX\_URL}\\
\textcolor{appendixdiffgreen}{+ \ \ || 'https://\textless BACKEND\textgreater.convex.cloud',}\\
\textcolor{appendixrulegray}{...}}
}

\vspace{0.65em}
{\footnotesize
\begin{tabular}{@{}p{0.26\linewidth}p{0.67\linewidth}@{}}
\textbf{Component} & Instruction + Script \\
\textbf{Operation} & Modify + Modify \\
\textbf{Intent} & \texttt{INTENT\_10}: Compatibility and interoperability \\
\textbf{Pattern Family} & \texttt{FAMILY\_08}: Executable workflow automation \\
\end{tabular}}
\end{minipage}
\hfill
{\color{appendixrulegray}\vrule width 0.4pt}
\hfill
\begin{minipage}[t]{0.485\textwidth}
\raggedright
\textbf{(b) Trajectory-derived change}\hfill{\small SearchQA $\cdot$ SkillOpt / GPT-5.2}\\[-0.25em]
{\color{appendixrulegray}\rule{\linewidth}{0.5pt}}
\vspace{0.45em}

\appendixdiffbox{%
{\bfseries\footnotesize SKILL.md}\\[0.25em]
{\ttfamily\fontsize{7.2}{8.8}\selectfont
\textcolor{appendixdiffgreen}{+ **Jeopardy/quiz dump guardrails:**}\\
\textcolor{appendixdiffgreen}{+ In lines formatted as}\\
\textcolor{appendixdiffgreen}{+ \textasciigrave CATEGORY | clue | \textless answer\textgreater. right: \textless contestant\textgreater\textasciigrave,}\\
\textcolor{appendixdiffgreen}{+ extract the text after the last \textasciigrave|\textasciigrave}\\
\textcolor{appendixdiffgreen}{+ and before \textasciigrave right:\textasciigrave\ or \textasciigrave wrong:\textasciigrave.}\\
\textcolor{appendixdiffgreen}{+ Do not output the contestant name.}\\[0.25em]
\textcolor{appendixdiffgreen}{+ For flashcard-style snippets, extract the short}\\
\textcolor{appendixdiffgreen}{+ standalone answer token at the end of the snippet.}\\
\textcolor{appendixrulegray}{...}}
}

\vspace{0.65em}
{\footnotesize
\begin{tabular}{@{}p{0.26\linewidth}p{0.67\linewidth}@{}}
\textbf{Component} & Instruction \\
\textbf{Operation} & Add \\
\textbf{Intent} & \texttt{INTENT\_04}: Evidence retrieval quality \\
\textbf{Pattern Family} & \texttt{FAMILY\_01}: Evidence retrieval and navigation \\
\end{tabular}}
\end{minipage}
\par
\caption{Representative skill-change units from public version histories and trajectory-driven evolution. The public update coordinates instruction and executable-code changes to preserve compatibility with an external service. The trajectory-derived update adds a local extraction rule in response to an observed SearchQA failure.}
\label{fig:representative-change-units}
\end{figure*}

\subsection{Broader Public-History Statistics}
\label{app:public-history-statistics}

Separate from the BFCL/SearchQA matched comparison, we survey 1,904 GitHub skills and 242 ClawHub skills across 12 task domains.
The survey contains 36,719 substantive adjacent-version updates from 2,146 public skills.
For each update, we identify the affected components and assign its dominant high-level intent.

\begin{figure}[t]
\centering
\includegraphics[width=\columnwidth]{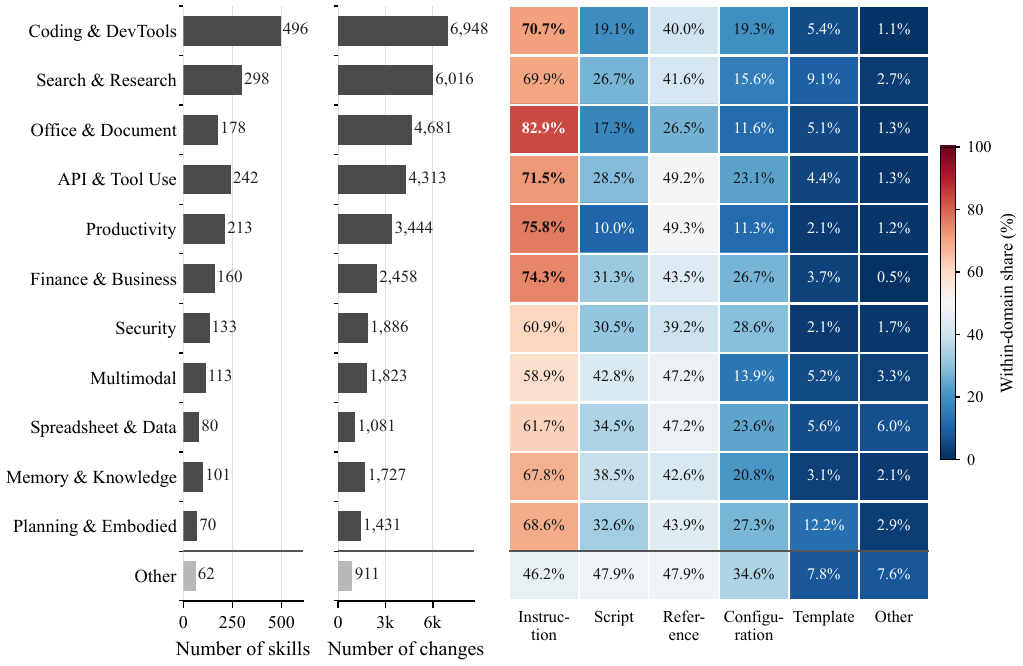}
\caption{Skill counts, substantive version updates, and affected-component shares across 12 task domains in the broader public-history survey. Component shares are non-exclusive because one update may affect several artifacts.}
\label{fig:finding1}
\end{figure}

Figure~\ref{fig:finding1} shows that public evolution routinely changes instructions, scripts, references, and configurations.
Instructions are the most frequently modified component in 11 of the 12 domains, with within-domain shares ranging from 58.9\% to 82.9\% in those domains.
Scripts, references, and configurations also occur throughout the domain set, capturing implementation and maintenance knowledge beyond instruction editing.

\subsection{Additional Analyses of Trajectory-Driven Evolution}
\label{app:trajectory-analyses}

\paragraph{Trajectory utility.}
Each trajectory is scored on six dimensions: execution integrity, feedback fidelity, trace observability, root-cause diagnosability, optimization actionability, and cross-task generalizability.
Each dimension receives an integer score $r_d\in\{0,1,2\}$, giving a total score $T(\tau)=\sum_{d=1}^{6}r_d\in[0,12]$ and normalized utility $U(\tau)=T(\tau)/12$.
High-utility trajectories satisfy $T(\tau)\geq 10$ and contain no hard flag, medium-utility trajectories satisfy $7\leq T(\tau)\leq 9$, and low-utility trajectories satisfy $T(\tau)\leq 6$.
Any non-infrastructure hard flag assigns the trajectory to the low-utility group.
Timeouts, connection errors, API errors, exceptions, empty traces, and missing traces are labeled \texttt{infra\_invalid} before utility classification.
For iteration $t$, trajectory utility is the mean normalized utility of the newly collected trajectories used to produce the update.
Update effectiveness is $\Delta\mathrm{Val}_t=\mathrm{Val}(S_t)-\mathrm{Val}(S_{t-1})$, the validation-score change produced by that iteration.

\begin{figure}[t]
\centering
\includegraphics[width=\columnwidth]{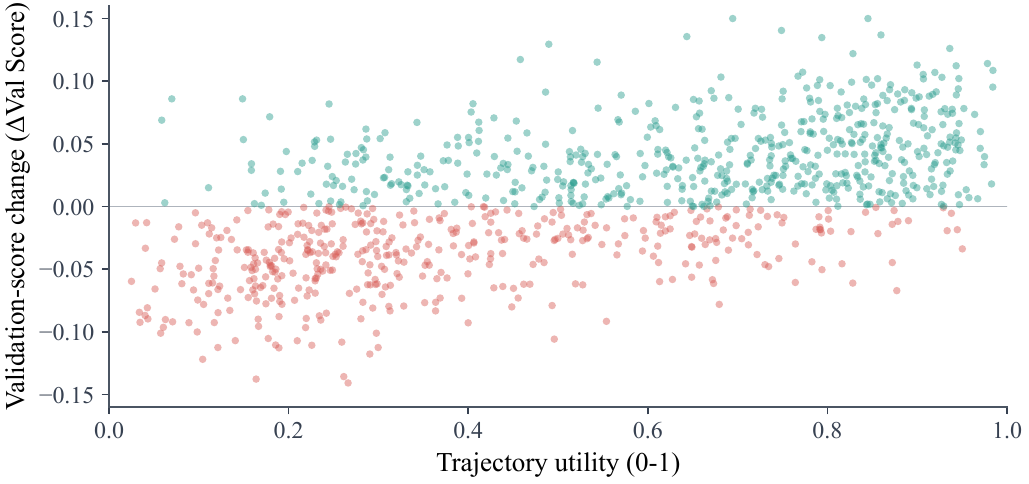}
\caption{Trajectory utility and validation-score change across skill-evolution iterations.}
\label{fig:finding3}
\end{figure}

Figure~\ref{fig:finding3} shows a positive association between trajectory utility and validation improvement.
Negative updates are concentrated among low-utility trajectories, while high-utility trajectories more often support positive validation changes.
The quality of the current rollout therefore directly affects the update signal available to trajectory-driven evolution.

\paragraph{Failure-mode identification.}
The judge records one primary failure mode, an optional secondary mode, confidence, rationale, and hard flags for each trajectory.
The ten canonical OfficeQA modes are document retrieval, table localization, row--column alignment, temporal scope, unit/scale/sign, arithmetic aggregation, answer extraction/format, unsupported or hallucinated evidence, tool execution, and other.
The normalizer first retains exact canonical labels, then maps markers for answer format, unsupported or hallucinated evidence, missing-document retrieval, table structure or columns, temporal scope, units/scales/signs, arithmetic/formulas/statistics, and tool execution to the corresponding modes.
Unmatched outputs are assigned to the other category.

\paragraph{Failure-mode coverage.}
For a set of trajectories, \texttt{unique\_modes} is the number $M$ of distinct primary failure modes and \texttt{mode\_coverage} is $M/|\mathcal{M}_{\mathrm{ref}}|$, where $\mathcal{M}_{\mathrm{ref}}$ is the reference mode set.
The normalized entropy of the mode distribution is
\[
H_{\mathrm{norm}}=-\frac{\sum_{m=1}^{M}p_m\log p_m}{\log M},
\]
with $H_{\mathrm{norm}}=0$ when $M\leq 1$.
The \texttt{hq\_diverse} condition covers at least four modes and selects at most two trajectories from each mode, while \texttt{hq\_narrow} contains trajectories from only one or two modes.
At the run level, we group observed coverage into low (1--3), medium (4--8), and high (9--10) bands.
We report the changes in training and test performance, their difference, and pass-to-fail regressions (P2F), defined as tasks that pass before evolution and fail afterward.

\begin{table}[t]
\centering
\small
\setlength{\tabcolsep}{5pt}
\begin{tabular}{lrrrr}
\toprule
Failure-mode coverage & $\Delta$Train $\uparrow$ & $\Delta$Test $\uparrow$ & Gap $\downarrow$ & P2F $\downarrow$ \\
\midrule
Low (1--3 modes) & 4.82 & 2.54 & 2.28 & 12.7 \\
Medium (4--8 modes) & 4.17 & 2.81 & 1.36 & 9.7 \\
High (9--10 modes) & 4.23 & 3.19 & 1.04 & 5.9 \\
\bottomrule
\end{tabular}
\caption{Training-test gaps and pass-to-fail regressions under different failure-mode coverage.}
\label{tab:finding4}
\end{table}

Table~\ref{tab:finding4} shows that broader failure-mode coverage improves test gains and reduces regressions.
From the low- to high-coverage group, the train--test gap decreases from 2.28 to 1.04 and the mean P2F count decreases from 12.7 to 5.9.
These results connect narrow trajectory evidence with overfitting during iterative skill evolution.

The matched study and the supplementary analyses motivate the two information sources used by \tool{}.
Public histories contribute recurring maintenance and implementation strategies, and current trajectories contribute task-specific execution evidence.
Adaptive selection and fusion combine these signals throughout skill evolution.

\FloatBarrier

\section{\tool{} Implementation Details}
\label{app:method-details}

\renewcommand{\dbltopfraction}{0.9}
\renewcommand{\textfraction}{0.05}

\begin{table}[t]
\centering
\small
\begin{tabular}{p{0.18\columnwidth}p{0.72\columnwidth}}
\toprule
Symbol & Meaning \\
\midrule
$u_j$ & update event
$\langle d_j,\{c_j\},o_j,i_j\rangle$ containing a raw diff segment,
affected components, semantic operation, and intent \\
$\mathcal{B}$ & version-change experience bank containing update patterns,
evolution insights, and domain insights \\
$S_t$ & target skill retained by the base evolver after online iteration $t$ \\
$\tau_t$ & task-execution trajectory produced with $S_t$ \\
$z_t$ & selection context containing target task, current skill, and available prior feedback \\
$\mathcal{X}_t$ & at most five experience entries selected from $\mathcal{B}$ \\
$p_t^{\mathrm{exp}}$ & aggregated external-experience guidance \\
$p_t^{\mathrm{self}}$ & trajectory-grounded proposal from the base evolver \\
$p_t^{\mathrm{enh}}$ & provenance-preserving fusion of the two proposal sources \\
$f_t^S,f_t^V,f_t$ & source, validation, and unified feedback \\
$\lambda_{\mathrm{exp}}^{(t)}$ & prompt-level reliance weight assigned to external
experience; $\lambda_{\mathrm{self}}^{(t)}=1-\lambda_{\mathrm{exp}}^{(t)}$ \\
\bottomrule
\end{tabular}
\caption{Core notation used in the implementation.}
\label{tab:method-notation}
\end{table}

\subsection{Offline Experience-Bank Construction}
\label{app:experience-bank}

Algorithms~\ref{alg:offline-distillation} and~\ref{alg:online-vce} detail the offline and online stages described in the main paper.
Adjacent versions are used because they preserve the local maintenance context and reduce ambiguity about which edits belong to one update step.
The method first abstracts raw diffs into events, then progressively removes repository-specific detail at the pattern, skill, and domain levels.
The source corpus contains 1,904 GitHub skills and 242 ClawHub skills across 12 task domains, as described in Section~\ref{app:motivation}.
For each evaluation benchmark, we manually select the five skills most closely aligned with its task and tool-use requirements and distill their adjacent-version changes into a benchmark-specific experience bank.

\paragraph{Grouping and abstraction.}
Events are grouped when their components, semantic operations, and intents support the same reusable strategy.
Duplicate or paraphrased events are merged when they recommend the same operation under the same applicability conditions, while events with different conditions or intended outcomes remain separate.
An update pattern states transferable operational guidance and cites its supporting event identifiers.
Skill-level insights synthesize compatible patterns from one skill, and domain-level insights require support across multiple skills or distinct concerns in the same domain.
Conflicting observations yield conditional guidance when their applicability conditions explain the difference; otherwise, they are omitted.
Each abstraction preserves backward links to its supporting events for verification against concrete diffs.

\paragraph{Deduplication and quality control.}
The distillation prompts enforce evidence-linked identifiers at every level and merge entries with equivalent guidance and applicability.
They remove repository names, exact paths, version numbers, benchmark answers, and one-off implementation details from reusable guidance.
Entries are retained only when they are evidence-grounded, actionable, transferable, and no broader than their supporting changes.

\begin{figure*}[t]
\begin{minipage}[t]{0.485\textwidth}
\vspace{0pt}
\captionof{algorithm}{Offline version-change experience distillation}
\label{alg:offline-distillation}
\begin{algorithmic}[1]
\REQUIRE Public skills with chronologically ordered versions
\ENSURE Experience bank $\mathcal{B}$
\STATE $\mathcal{B}\leftarrow\emptyset$
\FOR{each retained skill $r$}
    \STATE order versions $(v_1,\ldots,v_{m_r})$ chronologically
    \STATE $\mathcal{U}_r\leftarrow\emptyset$
    \FOR{$j=1$ to $m_r-1$}
        \STATE $D_j\leftarrow\mathrm{Diff}(v_j,v_{j+1})$
        \STATE remove ineligible mechanical changes from $D_j$
        \STATE segment $D_j$ into semantic change units
        \FOR{each change unit $g$}
            \STATE extract $u=\langle d,\{c\},o,i\rangle$
            \STATE validate schema, labels, and evidence traceability
            \STATE $\mathcal{U}_r\leftarrow\mathcal{U}_r\cup\{u\}$
        \ENDFOR
    \ENDFOR
    \STATE group compatible events in $\mathcal{U}_r$ and distill update patterns
    \STATE synthesize the patterns into skill-level evolution insights
    \STATE add validated patterns and insights to $\mathcal{B}$
\ENDFOR
\FOR{each task domain represented in $\mathcal{B}$}
    \STATE generalize recurring evolution insights into domain insights
    \STATE add validated domain insights to $\mathcal{B}$
\ENDFOR
\STATE deduplicate entries without removing distinct supporting evidence
\RETURN $\mathcal{B}$
\end{algorithmic}
\end{minipage}\hfill
\begin{minipage}[t]{0.485\textwidth}
\vspace{0pt}
\captionof{algorithm}{Online optimization with adaptive experience attention}
\label{alg:online-vce}
\begin{algorithmic}[1]
\REQUIRE {Initial skill $S_0$, target task $\mathcal{T}$, experience bank $\mathcal{B}$, and base evolver $\mathcal{E}$}
\ENSURE Skill returned by the native base-evolver loop
\STATE $\lambda_{\mathrm{exp}}^{(1)}\leftarrow0.5$,
$\lambda_{\mathrm{self}}^{(1)}\leftarrow0.5$
\STATE execute $S_0$ to obtain $\tau_0$
\FOR{$t=1$ to the iteration budget}
    \IF{$t=1$}
        \STATE $z_t\leftarrow\mathrm{Context}(\mathcal{T},S_{t-1})$
    \ELSE
        \STATE $z_t\leftarrow\mathrm{Context}(\mathcal{T},S_{t-1},f_{t-1})$
    \ENDIF
    \STATE $\mathcal{X}_t\leftarrow
    \mathrm{Select}_{\leq 5}(\mathcal{B},z_t)$
    \STATE $p_t^{\mathrm{exp}}\leftarrow\mathrm{Aggregate}(\mathcal{X}_t)$
    \STATE $p_t^{\mathrm{self}}\leftarrow
    \mathcal{E}.\mathrm{Propose}(S_{t-1},\tau_{t-1})$
    \STATE attach unique \texttt{EXP-k} and \texttt{SELF-k} identifiers
    \STATE $p_t^{\mathrm{enh}}\leftarrow
    \mathrm{Fuse}(p_t^{\mathrm{exp}},p_t^{\mathrm{self}};
    \lambda_{\mathrm{exp}}^{(t)},\lambda_{\mathrm{self}}^{(t)})$
    \STATE $\widehat{S}_t\leftarrow\mathcal{E}.\mathrm{Apply}(S_{t-1},p_t^{\mathrm{enh}})$
    \STATE $(S_t,o_t)\leftarrow\mathcal{E}.\mathrm{Retain}(S_{t-1},\widehat{S}_t)$
    \STATE align the tagged edits with $\mathrm{Diff}(S_{t-1},\widehat{S}_t)$
    \STATE compute $f_t^S$, obtain $f_t^V$ from $o_t$, and set $f_t=f_t^Vf_t^S$
    \STATE execute retained $S_t$ to obtain $\tau_t$
    \STATE $\lambda_{\mathrm{exp}}^{(t+1)}\leftarrow
    \mathrm{clip}(\lambda_{\mathrm{exp}}^{(t)}+0.1f_t,0.3,0.7)$
    \STATE $\lambda_{\mathrm{self}}^{(t+1)}\leftarrow
    1-\lambda_{\mathrm{exp}}^{(t+1)}$
\ENDFOR
\RETURN the skill selected by $\mathcal{E}$
\end{algorithmic}
\end{minipage}
\end{figure*}

\subsubsection{Distillation Prompt Definitions}
\label{app:distillation-prompts}

\paragraph{Update-event extraction.}
The extractor receives one diff chunk between adjacent skill versions and returns zero or more atomic semantic events.
Each event separates the exact diff evidence, affected components, semantic operation, and supported update intent.

\begin{figure*}[t]
\centering
\begin{minipage}{0.96\textwidth}
\begin{lstlisting}[style=appendixprompt,basicstyle=\ttfamily\fontsize{7}{8}\selectfont,framesep=5pt]
You extract structured update events from one supplied diff chunk between adjacent versions of an agent skill.

An update event is one atomic semantic change: one coherent operation on one or more closely related components for one supported intent. Split independent changes into separate events. Use only evidence present in the supplied diff.

Extraction rules:
- Include changes that alter skill behavior, capability, reliability, safety, tool use, validation, recovery, or task execution.
- Ignore formatting-only edits, generated artifacts, dependency-lock churn, path-only renames, and wording changes that preserve behavior.
- `raw_diff_segment` must contain only exact relevant lines from the supplied diff. Do not reconstruct or paraphrase those lines.
- `components` names the affected functional artifacts, such as instructions, scripts, references, configuration, metadata, or templates.
- `operation` states what changed. `intent` states the improvement purpose. Keep them distinct.
- Do not infer an intent that is not supported by the change. Omit that event instead of inventing rationale.
- Merge lines only when they implement the same atomic change.
- Return an empty list when the diff contains no meaningful semantic update.

Before returning, verify that every event is atomic, evidence-grounded, non-duplicative, and free of repository facts not present in the input.

Return JSON only:
{
  "events": [
    {
      "raw_diff_segment": "exact relevant diff lines",
      "components": ["affected component"],
      "operation": "what semantic change was made",
      "intent": "why the supported change improves the skill"
    }
  ]
}
\end{lstlisting}
\end{minipage}
\par
\caption{System prompt for update-event extraction.}
\label{fig:update-event-prompt}
\end{figure*}

\paragraph{Update-pattern distillation.}
The input contains one skill name and its structured update events.

\paragraph{Skill-level insight synthesis.}
The input contains the update patterns distilled from one skill history.

\begin{figure*}[t]
\centering
\begin{minipage}{0.96\textwidth}
\begin{lstlisting}[style=appendixprompt,basicstyle=\ttfamily\fontsize{7}{7.5}\selectfont,framesep=4pt]
You distill supplied update events from one skill's version history into reusable update patterns.

An update pattern is transferable guidance about how and why to evolve a skill. Group compatible events by component, operation, and intent, while preserving meaningful applicability boundaries.

Distillation rules:
- Every pattern must cite one or more supplied event IDs in `supporting_event_ids`.
- Cite only supplied event IDs. Never invent, rewrite, or omit the grounding IDs for a non-empty pattern.
- Merge duplicate or paraphrased patterns when they recommend the same operation under the same conditions.
- Keep separate patterns when their applicability or intended outcome differs.
- Remove repository names, exact paths, version numbers, benchmark-specific answers, and one-off implementation details.
- State constructive, operational guidance that can transfer to another skill.
- Reject observations that describe a change but do not provide a reusable evolution decision.
- Return an empty list when no supplied event supports a transferable pattern.

Before returning, verify that patterns are grounded, non-duplicate, transferable, actionable, and no broader than their supporting evidence.

Return JSON only:
{
  "patterns": [
    {
      "content": "reusable update pattern",
      "applicability": "conditions under which this pattern is useful",
      "supporting_event_ids": ["EVENT-id"]
    }
  ]
}
\end{lstlisting}
\end{minipage}
\par
\captionof{figure}{System prompt for update-pattern distillation.}
\label{fig:pattern-prompt}
\vspace{0.8em}

\begin{minipage}{0.96\textwidth}
\begin{lstlisting}[style=appendixprompt,basicstyle=\ttfamily\fontsize{7}{7.5}\selectfont,framesep=4pt]
You synthesize supplied update patterns from one skill into evolution insights.

An evolution insight is an actionable strategy for improving similar skills, not a summary of one repository change.

Synthesis rules:
- Prefer insights supported by multiple compatible patterns.
- A single pattern may support an insight only when it clearly expresses a general evolution strategy rather than an incidental fix.
- Every insight must cite only supplied pattern IDs in `supporting_pattern_ids`.
- Preserve applicability boundaries. Do not generalize beyond the situations represented by the supporting patterns.
- Merge overlapping insights when one actionable formulation covers both.
- Keep insights separate when they address different failure modes or applicability conditions.
- Exclude benchmark answers, repository details, unsupported causal claims, and purely descriptive observations.
- Return an empty list when no reliable skill-level insight is supported.

Before returning, verify grounding, applicability, actionability, non-duplication, and that incidental edits were not promoted into principles.

Return JSON only:
{
  "insights": [
    {
      "content": "skill-level evolution insight",
      "applicability": "when another skill should use this insight",
      "supporting_pattern_ids": ["PATTERN-id"]
    }
  ]
}
\end{lstlisting}
\end{minipage}
\par
\captionof{figure}{System prompt for skill-level evolution-insight synthesis.}
\label{fig:skill-insight-prompt}
\end{figure*}

\paragraph{Domain-level insight generalization.}
The input contains skill-level insights from multiple skills in one task domain.

\subsection{Online Optimization Algorithm}
\label{app:full-optimization}

Algorithm~\ref{alg:online-vce} gives the complete online recurrence.
\tool{} changes how the update proposal is constructed and leaves candidate evaluation, retention, checkpointing, and stopping to the native mechanism of the base evolver.

\subsection{Experience Selection and Aggregation}
\label{app:experience-selection}

The selector receives the benchmark, current skill, compact entries from the corresponding benchmark-specific bank, and the preceding unified feedback when one is available.
It evaluates each entry by task relevance, correspondence to an observable gap in the current skill, next-step actionability, applicability, redundancy, and consistency with the preceding feedback.
The selection budget is $K=5$; the output contains zero to five unique bank identifiers together with compact external guidance linked to the selected entries.
The selector prefers a small complementary set over redundant restatements and returns an empty selection when no supplied entry is sufficiently relevant and actionable.
When the selection is empty, fusion is skipped and the base evolver's self proposal proceeds unchanged.

\paragraph{Role of prior feedback.}
Task and skill compatibility remain the primary selection criteria.
Positive $f_{t-1}$ makes selection more receptive to external guidance, whereas negative feedback requires stronger applicability evidence.
The first iteration contains no prior feedback, so selection uses only the benchmark and current skill.
The source weights specify relative reliance on external and self-generated proposals.

\begin{figure*}[t]
\centering
\begin{minipage}{0.96\textwidth}
\begin{lstlisting}[style=appendixprompt,basicstyle=\ttfamily\fontsize{7}{8}\selectfont,framesep=5pt]
You generalize supplied evolution insights from multiple skills in the same task domain.

A domain insight is an operational optimization principle that transfers across skills or across clearly different skill concerns in the domain.

Generalization rules:
- Every insight must cite only supplied skill-insight IDs in `supporting_skill_insight_ids`.
- Require evidence across skills or clearly distinct concerns. Do not rename a single narrow skill insight as a domain principle.
- Preserve applicability conditions and operational actions.
- Merge duplicate principles with equivalent actions and applicability.
- When supplied insights conflict, keep conditional guidance only if the differing applicability conditions explain the conflict; otherwise omit the unsupported generalization.
- Exclude repository details, dataset instances, benchmark-specific answer knowledge, and unsupported claims.
- Return an empty list when cross-skill evidence is insufficient.

Before returning, verify cross-skill support, transferability, conflict handling, non-duplication, and valid supplied IDs.

Return JSON only:
{
  "insights": [
    {
      "content": "domain-level evolution insight",
      "applicability": "when a target skill should use this insight",
      "supporting_skill_insight_ids": ["SKILL-id"]
    }
  ]
}
\end{lstlisting}
\captionof{figure}{System prompt for domain-level evolution-insight generalization.}
\label{fig:domain-insight-prompt}
\end{minipage}
\vspace{0.8em}

\begin{minipage}{0.96\textwidth}
\begin{lstlisting}[style=appendixprompt,basicstyle=\ttfamily\fontsize{7}{8}\selectfont,framesep=5pt]
You select external version-change experiences for the current skill state.

Evaluate each supplied bank entry using:
- relevance to the benchmark and current task type;
- relevance to observable gaps or omissions in the current skill;
- actionability in the next optimization step;
- applicability conditions that match the current situation;
- redundancy with other selected experiences;
- consistency with the previous unified feedback.

Selection rules:
- Select no more than `selection_budget` supplied experience IDs.
- Select only supplied experience IDs.
- Selection may contain zero experiences when none is sufficiently relevant or actionable.
- Prefer a small, complementary set over redundant restatements.
- Positive previous feedback permits stronger reliance on relevant external experience.
- Negative previous feedback requires more conservative selection: demand stronger relevance and avoid repeating the previously emphasized strategy. It does not automatically ban all external experience.
- External guidance must summarize only selected entries and must not propose exact patch syntax, benchmark answers, or unsupported facts.
- Every guidance item must cite its selected source experience IDs.

Before returning, verify budget compliance, supplied IDs, non-redundancy, current-skill relevance, and that zero selection was used when evidence was weak.

Return JSON only:
{
  "selected_experience_ids": ["experience-id"],
  "external_guidance": [
    {
      "guidance": "compact transferable guidance",
      "source_experience_ids": ["experience-id"]
    }
  ],
  "reasoning": "brief selection rationale"
}
\end{lstlisting}
\end{minipage}
\par
\captionof{figure}{System prompt for feedback-conditioned experience selection and aggregation.}
\label{fig:selection-prompt}
\end{figure*}

\subsection{Attention Fusion and Skill Optimization}
\label{app:attention-fusion}
\label{app:fusion-prompts}

The source weights are prompt-level reliance instructions.
At initialization, $\lambda_{\mathrm{exp}}^{(1)}=\lambda_{\mathrm{self}}^{(1)}=0.5$.
They serve as relative preferences for incompatible recommendations, rather than output quotas or sampling probabilities.
Direct trajectory evidence remains authoritative: the fuser preserves useful self edits and uses external guidance to refine, constrain, complement, or replace them when the guidance identifies a concrete conflict or stronger applicable strategy.
At equal weights, the same evidence rule resolves conflicts.
Duplicate, contradictory, unsupported, non-actionable, and out-of-scope edits are removed.

Every external item receives an identifier \texttt{EXP-k}, and every self-proposal item receives an identifier \texttt{SELF-k}.
A fused edit contains all source identifiers that materially support it, and mixed provenance is used only when one edit genuinely combines both sources.
The fuser emits \texttt{replace}, \texttt{append}, \texttt{prepend}, or \texttt{delete} operations over the current skill and returns an empty edit list when neither source supports an applicable change.
The shared rewriter applies the selected operations to the complete skill document, preserves effective existing guidance and protected blocks, and consolidates overlapping instructions.

\paragraph{Full-skill rewrite.}
After fusion, the shared skill optimizer receives the current skill and selected revision suggestions and returns the complete rewritten skill document.

\begin{figure*}[t]
\centering
\begin{minipage}{0.96\textwidth}
\begin{lstlisting}[style=appendixprompt,basicstyle=\ttfamily\fontsize{7}{8}\selectfont,framesep=5pt]
You fuse an independently generated self proposal with selected external evolution guidance into candidate skill edits.

The attention weights are relative conflict preferences, not output quotas and not probabilities. They guide which source to prefer when both sources make incompatible recommendations. Do not manufacture edits to match a numerical ratio.

Fusion rules:
- Direct trajectory evidence supporting the self proposal remains authoritative. External guidance may refine, constrain, complement, or replace a self edit only when semantically justified.
- Preserve useful self edits even when `lambda_exp` is larger, unless relevant external guidance identifies a concrete conflict or stronger transferable strategy.
- Use external-only provenance for an edit derived only from external guidance, self-only provenance for an unchanged self edit, and mixed provenance only when an edit genuinely combines both sources.
- Copy provenance IDs exactly from the supplied inputs.
- Remove duplicate, contradictory, unsupported, non-actionable, and out-of-scope edits.
- Every output edit must be a valid patch operation using `replace`, `append`, `prepend`, or `delete`, with the required target and content fields.
- Return an empty edit list rather than inventing an unsupported change.

Before returning, verify valid patch operations, exact provenance, conflict handling, non-duplication, and that the attention weights were used only as relative conflict preferences.

Return JSON only:
{
  "reasoning": "brief fusion rationale",
  "edits": [
    {
      "op": "replace|append|prepend|delete",
      "target": "exact target when required",
      "content": "edit content when required",
      "provenance_ids": ["SELF-1", "EXP-1"]
    }
  ]
}
\end{lstlisting}
\end{minipage}
\par
\captionof{figure}{System prompt for adaptive-attention fusion.}
\label{fig:fusion-prompt}
\vspace{0.8em}

\begin{minipage}{0.96\textwidth}
\begin{lstlisting}[style=appendixprompt,basicstyle=\ttfamily\fontsize{7}{8}\selectfont,framesep=5pt]
You are an expert skill-document rewriter for an AI agent training system.

You will receive:
1. The current skill document
2. A selected set of revise_suggestions distilled from trajectory analysis

Your job is to rewrite the FULL target skill document so it incorporates the selected suggestions coherently.

Hard requirements:
1. Produce a complete standalone skill document, not a patch.
2. Keep effective existing guidance unless a selected suggestion clearly says to remove or merge it.
3. Prefer consolidation and clarity over making the document longer.
4. Do not hardcode benchmark-specific answers, entity names, file paths, or gold values.
5. Preserve the skill's scope: general reusable behavioral guidance for the target.
6. Do not modify content inside the protected slow-update block between <!-- SLOW_UPDATE_START --> and <!-- SLOW_UPDATE_END --> except to keep it intact.
7. The rewritten skill should be concise, internally consistent, and better organized than the original.

Respond ONLY with a valid JSON object:
{
  "reasoning": "<why this rewrite implements the selected suggestions well>",
  "change_summary": ["<short change 1>", "<short change 2>"],
  "new_skill": "<the full rewritten skill document>"
}
\end{lstlisting}
\end{minipage}
\par
\captionof{figure}{System prompt for full-skill rewriting.}
\label{fig:optimizer-prompt}
\end{figure*}

\subsection{Provenance Attribution and Adaptation Feedback}
\label{app:provenance-feedback}
\label{app:attribution-prompts}

Let $\mathcal{C}_t=\{c_{t,k}\}_{k=1}^{n_t}$ be the provenance-tagged candidate edits.
The attribution stage aligns each item with the optimizer-produced candidate diff $\Delta S_t=\mathrm{Diff}(\widehat{S}_t,S_{t-1})$.
It assigns $+1$ to a realized edit supported by external experience, $-1$ to a realized edit supported by the self proposal, and $0$ when an edit is unrealized, cannot be aligned reliably, or has mixed provenance.
We compute the source feedback as follows.
\[
    f_t^S=\frac{1}{n_t}\sum_{k=1}^{n_t}a_{t,k}.
\]
The denominator includes all candidate edits, so a proposal that is largely ignored has a source-feedback magnitude near zero.

The native base-evolver outcome supplies binary validation feedback:
\[
f_t^V =
\left\{
\begin{array}{ll}
+1, & \Delta_t^V>0,\\
-1, & \Delta_t^V\leq0,
\end{array}
\right.
\qquad
f_t=f_t^Vf_t^S.
\]
Consequently, an improving update increases reliance on the source that contributed more realized edits, while a non-improving update shifts reliance away from that source.
We apply the following attention update.
\[
\lambda_{\mathrm{exp}}^{(t+1)}
=\mathrm{clip}\!\left(\lambda_{\mathrm{exp}}^{(t)}+0.1f_t,0.3,0.7\right)
\]
keeps both evidence sources active.

\paragraph{Boundary cases.}
When $n_t=0$, source feedback is $f_t^S=0$ and the attention weights remain unchanged.
Mixed-provenance edits receive zero source attribution.
Partial implementations, ambiguous matches, contradicted changes, no-op changes, wording-only overlap, and unapplied edits are marked unrealized.
Attribution is computed on the optimizer-produced candidate diff; the base evolver's subsequent rejection is represented by negative validation feedback.

\subsection{Base-Evolver Interfaces}
\label{app:base-evolver-interface}

\tool{} inserts a thin proposal adapter between each base evolver's native proposal stage and its native skill-update stage.
The adapter exposes the current skill and self proposal to the common selector--fuser interface, attaches \texttt{SELF-k} identifiers, and converts the fused result back to the base evolver's expected format.
The rollout collection, diagnosis, candidate evaluation, retention, stopping, and checkpoint logic of the base evolver remain unchanged.

\paragraph{Adapter mappings.}
For EvoSkill, the adapter wraps the full-skill revision derived from failure trajectories as one \texttt{replace} or \texttt{append} self edit, fuses it with external guidance, materializes a complete \texttt{SKILL.md}, and returns it to the native frontier and retention loop.
For SkillClaw, it renders the current and proposed skills, wraps their change as a self edit, materializes the fused proposal, and returns it to the native validation and publication cycle.
For SkillOpt, it attaches provenance to the native patch distilled from scored rollouts, fuses it with external edits in the same patch schema, and passes it to the existing ranking, patch-application, and retention stages.
EvoSkill and SkillClaw therefore use complete skill documents at the adapter boundary, whereas SkillOpt operates directly on bounded patch edits.
Adapting a new base evolver requires proposal serialization, provenance attachment, and result materialization.

\subsection{End-to-End Evolution Case Study}
\label{app:end-to-end-case}

\begin{figure*}[t]
\centering
\begin{minipage}{0.96\textwidth}
\begin{lstlisting}[style=appendixprompt,basicstyle=\ttfamily\fontsize{7}{8}\selectfont,framesep=5pt]
You semantically match provenance-tagged ranked candidate edits to the actual skill diff produced by the optimizer.

Attribution rules:
- Evaluate each supplied `candidate_id` independently.
- Set `realized=true` only when the actual skill diff implements the candidate's core semantic change.
- Exact wording is not required when behavior and intent are equivalent.
- Partial implementations, ambiguous similarity, contradicted changes, no-op changes, wording-only overlap, and unselected or unapplied changes are `realized=false`.
- Evidence must briefly identify the relevant actual skill diff content.
- If the actual diff provides insufficient evidence, use `realized=false`.
- Do not decide whether external or self experience was beneficial and do not assign source scores. Deterministic code calculates source feedback from provenance IDs.
- Return one match record for each supplied candidate ID and do not invent IDs.

Before returning, verify candidate coverage, conservative semantic matching, actual-diff evidence, valid supplied IDs, and boolean `realized` values.

Return JSON only:
{
  "matches": [
    {
      "candidate_id": "CAND-1",
      "realized": true,
      "evidence": "brief reference to the matching actual change"
    }
  ]
}
\end{lstlisting}
\captionof{figure}{System prompt for semantic realization matching.}
\label{fig:attribution-prompt}
\end{minipage}
\vspace{0.8em}

\begin{minipage}{\textwidth}
\centering
\includegraphics[width=\textwidth]{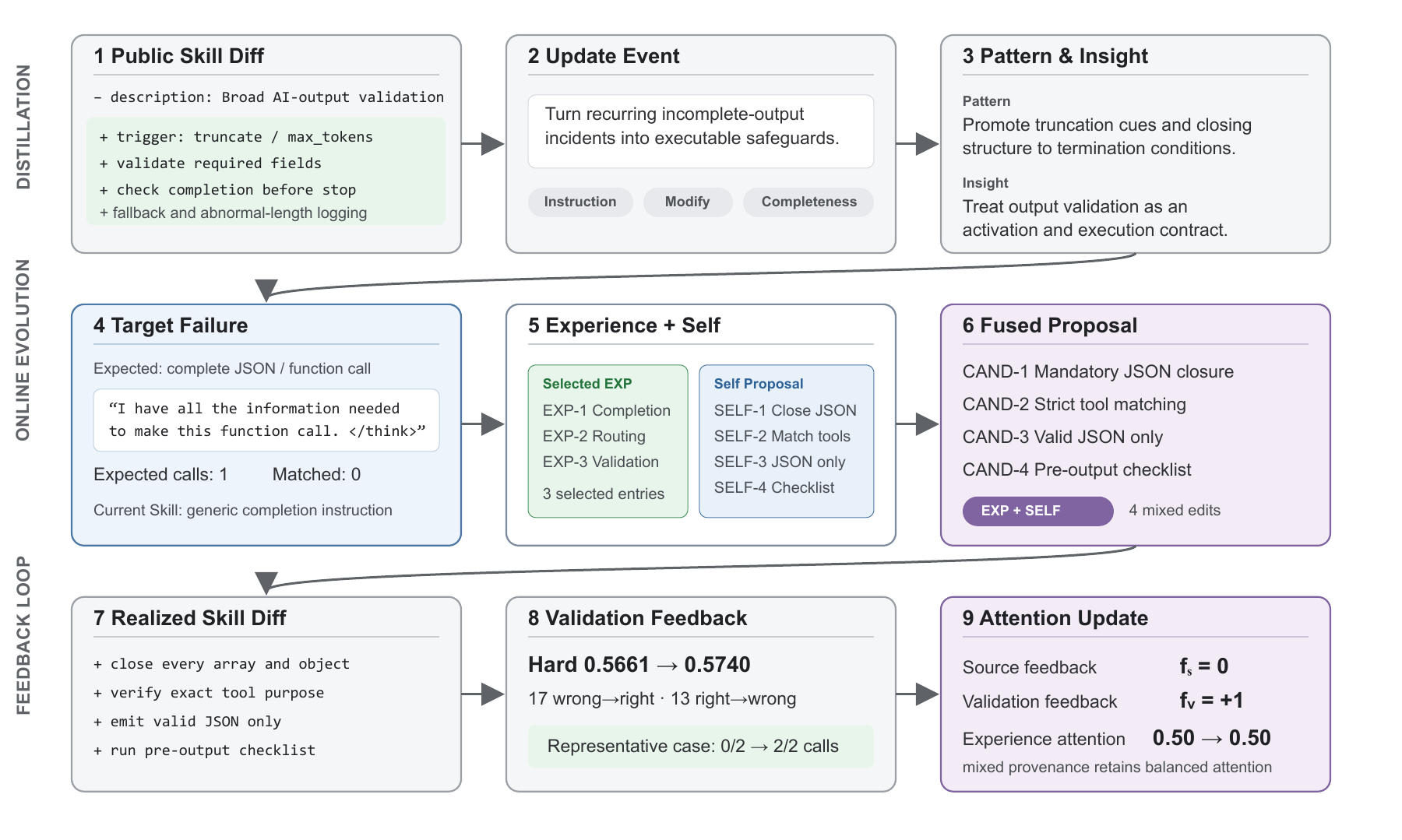}
\captionof{figure}{End-to-end evolution case.
A public version change addressing incomplete structured outputs is distilled into reusable experience.
The selected experience and self proposal are fused into four mixed-provenance edits, all of which are realized in the accepted Skill update.
The hard score increases from 0.5661 to 0.5740, while mixed provenance yields zero source feedback and retains balanced experience attention.}
\label{fig:end-to-end-case}
\end{minipage}
\end{figure*}

Figure~\ref{fig:end-to-end-case} follows one evolution trace from a public version change to feedback-driven attention adaptation.
The source update converts recurring incomplete-output incidents into explicit activation cues, completeness checks, and recovery rules.
Its update event, pattern, and insight provide reusable guidance for skills that produce structured outputs.

The target trajectory completes its analysis but stops before emitting the required function call.
The selector retrieves three complementary experiences, while the base evolver proposes four edits covering JSON closure, tool matching, output formatting, and pre-output validation.
The fuser preserves both sources in four mixed-provenance edits, and all four edits appear in the accepted Skill update.

Across 507 paired validation examples, the hard score increases from 0.5661 to 0.5740, with 17 wrong-to-right and 13 right-to-wrong transitions.
The representative parallel-call example improves from zero matched calls out of two to two matched calls out of two.
All realized edits have mixed provenance, giving $f_t^S=0$, $f_t^V=+1$, and $f_t=0$; the next-round weights therefore remain $\lambda_{\mathrm{exp}}=\lambda_{\mathrm{self}}=0.5$.

\FloatBarrier

\FloatBarrier

\section{Additional Experimental Results and Analyses}
\label{app:additional-experiments}

Unless otherwise specified, all supplementary experiments follow the main-paper protocol over SearchQA, OfficeQA, ALFWorld, Spreadsheet, and BFCL-v4 with Qwen3.5-27B, GPT-5.2, DeepSeek-v3.2, and Claude Sonnet 5.
Each configuration is independently run three times, and paired comparisons use the same benchmark instances, initial skill, evolution budget, and target model.
The compared methods are No Skill, LLM Skill, EvoSkill, SkillClaw, SkillOpt, and the VCE-enhanced variants of the three iterative self-evolvers.

We use GPT-5.5 for the LLM-based annotation and analysis in the motivation study.
For experience distillation and fusion in \tool{}, we use the corresponding target model in each experimental configuration.
All large language models are accessed through their APIs.
All experiments are conducted on a server equipped with an Intel Core i7-10700 CPU, an NVIDIA TITAN RTX GPU, and 32\,GB of RAM.

\subsection{Ablation Results Across Base Self-Evolvers}
\label{app:rq2-full}

\begin{figure}[t]
    \centering
    \includegraphics[width=\columnwidth]{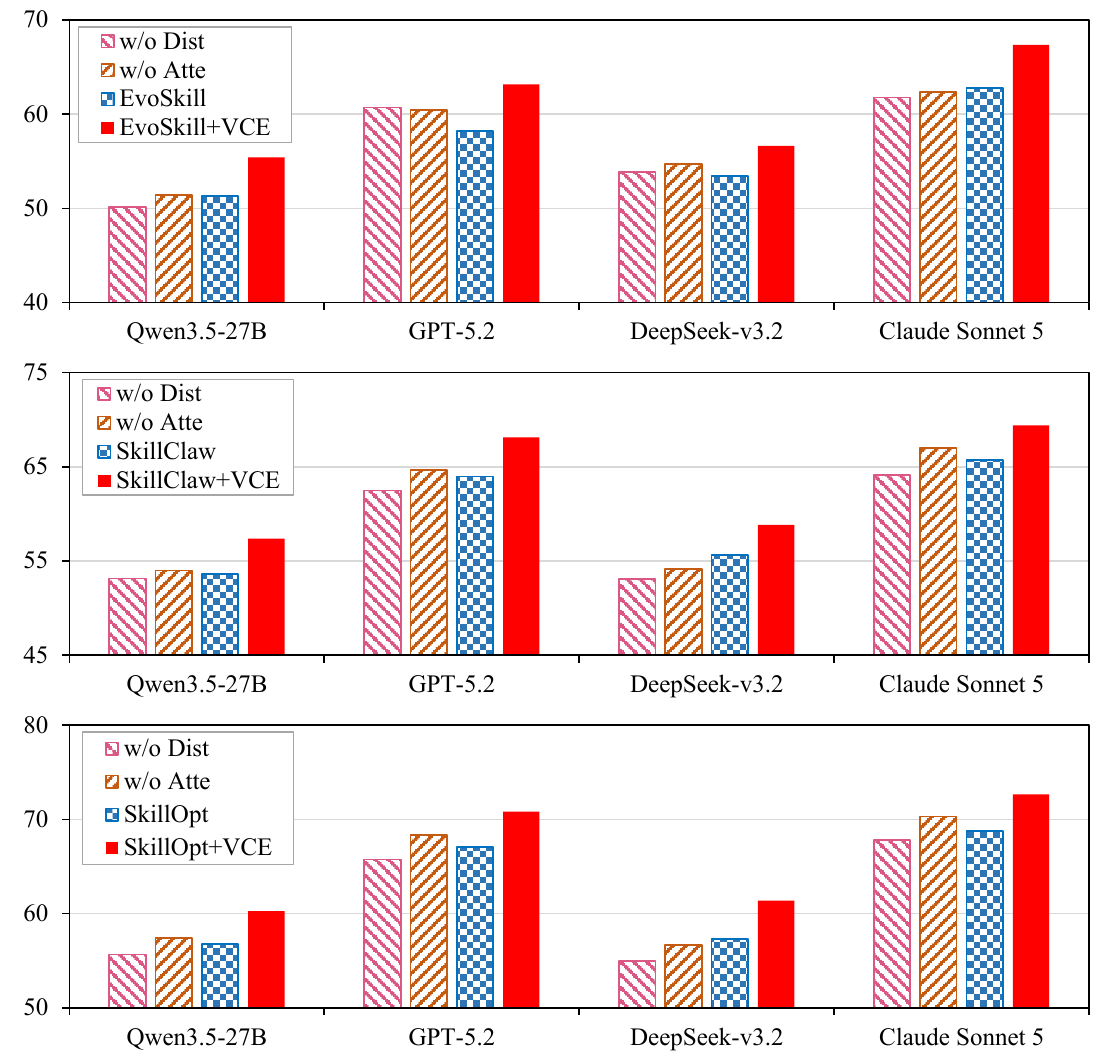}
    \caption{Ablation results averaged over all five benchmarks.
    Each panel uses EvoSkill, SkillClaw, or SkillOpt as the base self-evolver.
    \emph{w/o Dist} replaces distilled experience with raw version changes, and \emph{w/o Atte} assigns equal weights to external experience and self-generated proposals.
    Higher is better.}
    \label{fig:app-ablation}
\end{figure}

\paragraph{Experimental setup.}
We evaluate experience distillation and adaptive experience attention with EvoSkill, SkillClaw, and SkillOpt as the base self-evolvers.
For each base evolver, \emph{w/o Dist} replaces distilled experience with raw version-change records, while \emph{w/o Atte} replaces adaptive attention with fixed equal weights for external experience and self-generated proposals.
Figure~\ref{fig:app-ablation} reports the average performance across all five benchmarks for four agent models.

\paragraph{Results.}
The complete \tool{} achieves the highest average score for every base evolver and agent model, covering all twelve base-evolver--model combinations.
Removing experience distillation produces the larger degradation in most settings, indicating that distilled update patterns and evolution insights transfer more effectively than raw version-change records.
Removing adaptive experience attention also consistently reduces performance relative to the complete method, although fixed equal weighting usually retains part of the improvement over the corresponding base evolver.
These trends remain consistent across EvoSkill, SkillClaw, and SkillOpt and across Qwen3.5-27B, GPT-5.2, DeepSeek-v3.2, and Claude Sonnet 5.

\paragraph{Conclusion.}
Experience distillation and adaptive experience attention provide complementary improvements to \tool{}.
Distillation converts concrete version changes into reusable evolution knowledge, while adaptive attention controls the contribution of external experience during self-evolution.
Their consistent gains across three base self-evolvers demonstrate that both components generalize across different self-evolution frameworks.

\subsection{Hyperparameter and Attention Sensitivity}
\label{app:sensitivity}

\paragraph{Experimental setup.}
We evaluate the sensitivity of \tool{} to the experience-selection budget $K$, attention step size $\eta$, and attention clipping bounds.
All experiments use SkillOpt as the base self-evolver and Qwen3.5-27B as the agent model.
We vary one hyperparameter at a time while fixing the remaining settings to $K=5$, $\eta=0.1$, initial source weights of $0.5/0.5$, and clipping bounds of $[0.3,0.7]$.
Table~\ref{tab:sensitivity-matrix} reports the mean performance on each benchmark.

\paragraph{Results.}
The default configuration achieves the highest score on every benchmark for all three hyperparameters.
For the selection budget, $K=4$ performs within $0.02$--$0.11$ points of $K=5$, whereas increasing the budget to $K=7$ reduces performance by $1.11$--$1.88$ points.
The attention step size has the strongest effect: $\eta=0.1$ consistently performs best, while $\eta=0.2$ decreases performance by $2.28$--$3.88$ points across the five benchmarks.
The clipping bounds are comparatively stable around the default configuration.
The intervals $[0.4,0.6]$ and $[0.2,0.8]$ remain within $0.29$ points of the best result, whereas the wider interval $[0.1,0.9]$ produces a consistent reduction of $0.71$--$1.37$ points.

\paragraph{Conclusion.}
The results support $K=5$, $\eta=0.1$, and $[0.3,0.7]$ as the default configuration across all five benchmarks.
A moderate selection budget provides sufficient experience coverage without introducing excessive weakly relevant candidates.
The attention step size requires tighter calibration because it directly controls the rate of source-weight adaptation.
Moderately constrained clipping bounds maintain stable attention updates, whereas overly permissive bounds reduce performance.
Importantly, every evaluated configuration outperforms the corresponding SkillOpt baseline on all five benchmarks.
The hyperparameters affect the magnitude of improvement, and \tool{} retains a consistent advantage over SkillOpt throughout the evaluated ranges.

\begin{table*}[!t]
\centering
\scriptsize
\setlength{\tabcolsep}{3.2pt}
\begin{tabular}{@{}lllrrrrrrr@{}}
\toprule
Source Model & Target Model & Source Skill & SearchQA & OfficeQA & ALFWorld & Spreadsheet & BFCL-v4 & Avg. & Avg. Gain \\
\midrule

\multirow[c]{6}{*}{Qwen3.5-27B}
& \multirow[c]{2}{*}{GPT-5.2} & SkillOpt & 78.01 & 47.47 & 79.16 & 52.11 & 64.10 & 64.17 & -- \\
& & \textbf{SkillOpt+\tool{}} & \textbf{83.24} & \textbf{55.34} & \textbf{86.05} & \textbf{57.89} & \textbf{68.54} & \textbf{70.21} & \textbf{6.04} \\
& \multirow[c]{2}{*}{DeepSeek-v3.2} & SkillOpt & 71.16 & 44.94 & 69.60 & 38.16 & 62.10 & 57.19 & -- \\
& & \textbf{SkillOpt+\tool{}} & \textbf{78.12} & \textbf{49.53} & \textbf{75.11} & \textbf{43.94} & \textbf{65.48} & \textbf{62.44} & \textbf{5.25} \\
& \multirow[c]{2}{*}{Claude Sonnet 5} & SkillOpt & 81.10 & 53.03 & 83.69 & 56.01 & 66.42 & 68.05 & -- \\
& & \textbf{SkillOpt+\tool{}} & \textbf{87.37} & \textbf{58.14} & \textbf{86.80} & \textbf{61.73} & \textbf{71.97} & \textbf{73.20} & \textbf{5.15} \\

\midrule
\multirow[c]{6}{*}{GPT-5.2}
& \multirow[c]{2}{*}{Qwen3.5-27B} & SkillOpt & 75.81 & 47.66 & 61.24 & 50.62 & 56.36 & 58.34 & -- \\
& & \textbf{SkillOpt+\tool{}} & \textbf{82.04} & \textbf{53.59} & \textbf{66.49} & \textbf{56.46} & \textbf{60.89} & \textbf{63.89} & \textbf{5.56} \\
& \multirow[c]{2}{*}{DeepSeek-v3.2} & SkillOpt & 72.70 & 46.48 & 71.14 & 39.70 & 63.64 & 58.73 & -- \\
& & \textbf{SkillOpt+\tool{}} & \textbf{79.74} & \textbf{51.15} & \textbf{76.73} & \textbf{45.56} & \textbf{67.10} & \textbf{64.06} & \textbf{5.32} \\
& \multirow[c]{2}{*}{Claude Sonnet 5} & SkillOpt & 82.64 & 54.57 & 85.23 & 57.55 & 67.96 & 69.59 & -- \\
& & \textbf{SkillOpt+\tool{}} & \textbf{88.99} & \textbf{59.76} & \textbf{88.42} & \textbf{63.35} & \textbf{73.59} & \textbf{74.82} & \textbf{5.23} \\

\midrule
\multirow[c]{6}{*}{DeepSeek-v3.2}
& \multirow[c]{2}{*}{Qwen3.5-27B} & SkillOpt & 74.35 & 46.20 & 59.78 & 49.16 & 54.90 & 56.88 & -- \\
& & \textbf{SkillOpt+\tool{}} & \textbf{80.65} & \textbf{52.21} & \textbf{65.10} & \textbf{55.07} & \textbf{59.51} & \textbf{62.51} & \textbf{5.63} \\
& \multirow[c]{2}{*}{GPT-5.2} & SkillOpt & 78.09 & 47.55 & 79.24 & 52.19 & 64.18 & 64.25 & -- \\
& & \textbf{SkillOpt+\tool{}} & \textbf{83.47} & \textbf{55.57} & \textbf{86.28} & \textbf{58.12} & \textbf{68.77} & \textbf{70.44} & \textbf{6.19} \\
& \multirow[c]{2}{*}{Claude Sonnet 5} & SkillOpt & 81.18 & 53.11 & 83.77 & 56.09 & 66.50 & 68.13 & -- \\
& & \textbf{SkillOpt+\tool{}} & \textbf{87.60} & \textbf{58.37} & \textbf{87.03} & \textbf{61.96} & \textbf{72.20} & \textbf{73.43} & \textbf{5.30} \\

\midrule
\multirow[c]{6}{*}{Claude Sonnet 5}
& \multirow[c]{2}{*}{Qwen3.5-27B} & SkillOpt & 76.07 & 47.92 & 61.50 & 50.88 & 56.62 & 58.60 & -- \\
& & \textbf{SkillOpt+\tool{}} & \textbf{82.32} & \textbf{53.88} & \textbf{66.77} & \textbf{56.74} & \textbf{61.18} & \textbf{64.18} & \textbf{5.58} \\
& \multirow[c]{2}{*}{GPT-5.2} & SkillOpt & 79.81 & 49.27 & 80.96 & 53.91 & 65.90 & 65.97 & -- \\
& & \textbf{SkillOpt+\tool{}} & \textbf{85.14} & \textbf{57.24} & \textbf{87.95} & \textbf{59.79} & \textbf{70.44} & \textbf{72.11} & \textbf{6.14} \\
& \multirow[c]{2}{*}{DeepSeek-v3.2} & SkillOpt & 72.96 & 46.74 & 71.40 & 39.96 & 63.90 & 58.99 & -- \\
& & \textbf{SkillOpt+\tool{}} & \textbf{80.02} & \textbf{51.43} & \textbf{77.01} & \textbf{45.84} & \textbf{67.38} & \textbf{64.34} & \textbf{5.34} \\

\bottomrule
\end{tabular}
\caption{Complete cross-model transfer results over all 12 directed source--target model pairs.
Each source-trained skill is directly applied to the target model without further evolution.
Avg. reports the mean performance across the five benchmarks, and Avg. Gain reports the mean improvement of SkillOpt+\tool{} over SkillOpt computed from the unrounded results.
Bold indicates the higher result within each source--target pair.}
\label{tab:reported-transfer-means}
\end{table*}

\begin{table}[t]
\centering
\scriptsize
\setlength{\tabcolsep}{2.2pt}
\begin{tabular}{@{}llccccc@{}}
\toprule
Parameter & Value & SearchQA & OfficeQA & ALFWorld & Spreadsheet & BFCL-v4 \\
\midrule
\multirow[c]{5}{*}{$K$} & 3 & 76.24 & 47.54 & 61.30 & 51.39 & 55.98 \\
& 4 & 78.75 & 49.34 & 63.38 & 52.12 & 57.42 \\
& 5 & \textbf{78.86} & \textbf{49.42} & \textbf{63.43} & \textbf{52.14} & \textbf{57.52} \\
& 6 & 78.03 & 48.82 & 62.76 & 51.59 & 57.03 \\
& 7 & 76.98 & 48.07 & 61.90 & 50.89 & 56.41 \\
\midrule
\multirow[c]{5}{*}{$\eta$} & 0.01 & 75.16 & 46.96 & 61.42 & 50.67 & 56.34 \\
& 0.05 & 76.91 & 48.02 & 61.85 & 50.84 & 56.37 \\
& 0.10 & \textbf{78.86} & \textbf{49.42} & \textbf{63.43} & \textbf{52.14} & \textbf{57.52} \\
& 0.15 & 76.16 & 47.48 & 61.24 & 51.34 & 56.93 \\
& 0.20 & 74.98 & 46.83 & 60.28 & 49.55 & 55.24 \\
\midrule
\multirow[c]{4}{*}{Clipping} & $[0.4,0.6]$ & 78.61 & 49.14 & 63.23 & 51.97 & 57.37 \\
& $[0.3,0.7]$ & \textbf{78.86} & \textbf{49.42} & \textbf{63.43} & \textbf{52.14} & \textbf{57.52} \\
& $[0.2,0.8]$ & 78.57 & 49.21 & 63.41 & 52.05 & 57.48 \\
& $[0.1,0.9]$ & 77.49 & 48.54 & 62.62 & 51.43 & 56.71 \\
\bottomrule
\end{tabular}
\caption{Hyperparameter sensitivity of SkillOpt+\tool{} with Qwen3.5-27B. Each row reports mean performance on five benchmarks while varying one hyperparameter and fixing the remaining settings to their defaults. Bold indicates the best score within each hyperparameter group.}
\label{tab:sensitivity-matrix}
\end{table}

\subsection{Complete Cross-Model Transfer Results}
\label{app:rq3-full}

\paragraph{Experimental setup.}
We evaluate cross-model transfer over all ordered pairs of Qwen3.5-27B, GPT-5.2, DeepSeek-v3.2, and Claude Sonnet 5.
For each source model, SkillOpt and SkillOpt+\tool{} independently evolve a skill, which is then directly applied to each of the other three target models without further evolution.
This design yields 12 directed source--target pairs evaluated on five benchmarks.
The average gain is computed as the mean difference between SkillOpt+\tool{} and SkillOpt across the five benchmarks.

\paragraph{Results.}
SkillOpt+\tool{} outperforms SkillOpt on every benchmark for all 12 directed source--target pairs, yielding positive improvements in all 60 benchmark-level comparisons.
The average gains range from 5.15 to 6.19 points, with an overall mean improvement of 5.56 points.
Using Qwen3.5-27B, GPT-5.2, DeepSeek-v3.2, and Claude Sonnet 5 as the source yields average gains of 5.48, 5.37, 5.71, and 5.69 points, respectively.
The small variation across source models shows that the transfer improvement is preserved across different source-model capabilities.
Among the target models, GPT-5.2 receives the largest mean gain of 6.12 points, and every target model improves by at least 5.23 points on average.

\paragraph{Conclusion.}
The complete transfer results demonstrate that version-change experience improves the cross-model portability of evolved skills.
The improvement holds for every source model, target model, and benchmark evaluated in this study.
Because the transferred skills receive no target-side evolution, the consistent gains reflect the stronger transferable guidance produced by \tool{}.

\FloatBarrier

\section{Human Audit of LLM-based Decisions}
\label{sec:human-audit}

\paragraph{Audit Scope and Sampling.}
We conduct a human audit of six types of LLM-based semantic decisions: Motivation-study Taxonomy, Skill-Change Coding, Change Abstraction, Experience Generation, Experience Selection, and Source Attribution.

For Motivation-study Taxonomy, we sample 200 category--example instances from the frozen taxonomy and its calibration records.
The samples are balanced across the two task domains and stratified across the component, intent, and pattern dimensions.
Each instance contains one category definition and one supporting calibration example, allowing annotators to assess both the semantic validity of the category and whether the example supports it.

For Skill-Change Coding, we sample 200 annotated skill-change units.
The sample is balanced across the two task domains and two evolution sources, with 50 instances from each domain--source combination.
Each instance contains the paired skill states, the complete proposed segmentation as context, and one highlighted skill-change unit with its assigned component, intent, and pattern labels.
The audit target is the highlighted unit rather than every other unit in the complete segmentation.
Whenever possible, sampled units are drawn from different public version pairs or trajectory-evolution iterations to reduce sample dependence.
In addition, every category reported as public-only or trajectory-only is represented by at least one audited unit.

For each of the remaining four decision types, we sample 200 instances from the complete experimental records.
The samples are stratified across tasks, target skills, optimization iterations, and base evolvers to avoid over-representing a particular experimental setting.
For Experience Generation, the 200 samples consist of 67 update patterns, 67 evolution insights, and 66 domain insights.
Whenever possible, instances derived from different version pairs or optimization iterations are selected to reduce sample dependence.

%
%
%

\paragraph{Annotation Process.}
Two annotators familiar with agent skills independently inspect each instance.
The generating model, optimization method, and experimental condition are anonymized, while all evidence required for judgment is retained.
For Motivation-study Taxonomy, the explicit source identities of the supporting calibration examples are hidden.
For Skill-Change Coding, annotators are not informed whether an instance comes from the public or trajectory source.

Each annotator assigns a binary label, \emph{Accept} or \emph{Reject}, and provides a brief reason for rejected instances.
An output is accepted only when all corresponding criteria specified below are satisfied.
Disagreements are resolved by a third annotator.

Before the formal audit, the annotators complete a pilot round of 20 additional instances for each decision type.
These pilot instances are excluded from the final results.
If the inter-annotator agreement in a pilot round is below $\kappa=0.7$, the corresponding annotation guidelines are clarified and the pilot round is repeated.

%
%

\paragraph{Annotation Criteria.}
For \emph{Motivation-study Taxonomy}, annotators inspect a category definition and one of its supporting calibration examples.
The instance is accepted if the category represents a coherent evolution concept, is semantically distinguishable from other categories at the same level, is applicable independently of the evolution source, and is correctly supported by the given example.

For \emph{Skill-Change Coding}, annotators inspect the paired skill states, the complete segmentation as context, and one highlighted skill-change unit.
The instance is accepted only if the highlighted unit:
(1) groups edits that implement the same primary intent through the same modification pattern;
(2) includes all semantically coupled edits required to implement that change;
(3) excludes unrelated edits; and
(4) assigns component, intent, and pattern labels that are semantically correct and supported by the underlying skill change.
Errors in non-highlighted units do not affect the judgment of the current audit instance.

For \emph{Change Abstraction}, annotators inspect a raw diff and its generated update event.
The event is accepted if the raw diff segment corresponds to the actual change and its component, operation, and intent are all semantically correct and directly supported by the diff.

For \emph{Experience Generation}, annotators inspect an experience-bank entry together with its supporting lower-level items.
An entry is accepted if it faithfully summarizes its supporting items, conforms to the intended abstraction level, contains no unsupported claims, and provides reusable guidance for skill optimization.

For \emph{Experience Selection}, annotators inspect the selection context $z_t$ and one selected experience.
The selection is accepted if the experience is relevant to both the current task and the current skill state and provides applicable guidance for the current optimization step.

For \emph{Source Attribution}, annotators inspect a provenance-tagged candidate edit, the realized skill diff, and the predicted attribution.
The attribution is accepted if the candidate edit is correctly matched to the realized change and its label follows our definition: $+1$ for a realized externally supported edit, $-1$ for a realized self-supported edit, and $0$ for an unrealized, uncertain, or mixed-provenance edit.

%
%
%
%
%

\paragraph{Metrics.}
We report two common metrics for all six decision types.
First, the Human Acceptance Rate (HAR) is the proportion of outputs accepted after adjudication:
\begin{equation}
    \mathrm{HAR}
    =
    \frac{1}{N}
    \sum_{i=1}^{N}
    \mathbf{1}
    \left[
        \widetilde{y}_i=\mathrm{Accept}
    \right],
    \qquad N=200,
    \label{eq:human-acceptance-rate}
\end{equation}
where $\widetilde{y}_i$ is the final adjudicated label.

Second, we report Cohen's $\kappa$ between the two initial annotators to quantify the consistency of the annotation criteria.
HAR is computed from the final adjudicated labels, whereas Cohen's $\kappa$ is computed from the two annotators' independent labels before adjudication.

%

\begin{table}[t]
    \centering
    \small
    \setlength{\tabcolsep}{3.5pt}
    \caption{
        Human audit results for the LLM-based semantic decisions.
        HAR is computed from the final adjudicated labels, whereas
        $\kappa$ measures agreement between the two initial annotators
        before adjudication.
    }
    \label{tab:human-audit}
    \begin{tabular}{lccc}
        \toprule
        Decision Type
        & Samples
        & HAR (\%) $\uparrow$
        & $\kappa$ $\uparrow$ \\
        \midrule
        \multicolumn{4}{l}{\textit{Motivation study}} \\
        Motivation-study Taxonomy
        & 200 & 94.0 & 0.84 \\
        Skill-Change Coding
        & 200 & 100.0 & 0.84 \\
        \midrule
        \multicolumn{4}{l}{\textit{\tool{} method}} \\
        Change Abstraction
        & 200 & 92.5 & 0.82 \\
        Experience Generation
        & 200 & 90.5 & 0.78 \\
        Experience Selection
        & 200 & 86.5 & 0.75 \\
        Source Attribution
        & 200 & 90.5 & 0.80 \\
        \bottomrule
    \end{tabular}
\end{table}

\paragraph{Audit Results and Summary.}
As shown in Table~\ref{tab:human-audit}, the six audited decision types achieve HAR values ranging from 86.5\% to 100.0\%, with Cohen's $\kappa$ values between 0.75 and 0.84.
The two decisions used in the motivation study show strong agreement with human assessment.
Motivation-study Taxonomy achieves an HAR of 94.0\% and a $\kappa$ of 0.84, while all audited Skill-Change Coding instances are accepted after adjudication, with a pre-adjudication $\kappa$ of 0.84.
These results support the reliability of the taxonomy construction and skill-change annotations underlying the motivation-study findings.

The LLM-based decisions within \tool{} also remain consistently aligned with human judgments.
Change Abstraction, Experience Generation, and Source Attribution achieve HAR values above 90\%, indicating that most generated update events and hierarchical experience entries are faithful to their supporting changes and that most provenance assignments are consistent with the realized skill updates.
Experience Selection obtains the lowest HAR, at 86.5\%, which is consistent with the more context-dependent nature of determining whether an experience is applicable to the current task and skill state.
Nevertheless, its $\kappa$ of 0.75 indicates that the corresponding assessment criteria remain reasonably consistent across annotators.
Source Attribution achieves an HAR of 90.5\% and a $\kappa$ of 0.80, providing additional support for its use as the feedback signal in adaptive experience attention.

Overall, the high acceptance rates and consistent inter-annotator agreement provide empirical evidence that the LLM-based semantic decisions used in both the motivation study and \tool{} are sufficiently reliable for the analyses and optimization procedure.
These results do not imply that the LLM judgments are error-free; rather, they show that the outputs are generally supported by human assessment under the defined audit criteria.

%
%


\end{document}